\documentstyle[aaspp4,12pt,amsmath]{article}
\def \sun {$_{\scriptscriptstyle\odot}$}

\lefthead{FRYER, HOLZ, \& HUGHES} 

\righthead{Gravitational Wave Emission From Collapse}

\begin{document} 

\begin{center}

\end{center}

\vspace{1.cm}

\title{Gravitational Wave Emission From Core-Collapse of Massive Stars} 
\author{Chris L. Fryer} 
\affil{Theoretical Astrophysics, 
Los Alamos National Laboratories, \\ Los Alamos, NM 87544}
\authoremail{clf@t6-serv.lanl.gov}

\author{Daniel E. Holz and Scott A. Hughes}
\affil{Institute for Theoretical Physics, University of California, 
Santa Barbara, CA 93106}
\authoremail{deholz@itp.ucsb.edu}
\authoremail{hughes@itp.ucsb.edu}

\begin{abstract}

We derive estimates for the characteristics of gravitational radiation
from stellar collapse, using recent models of the core-collapse of
Chandrasekhar mass white dwarfs (accretion induced collapse),
core-collapse supernovae and collapsars, and the collapse of very
massive stars ($\gtrsim 300$\,M\sun\/).  We study gravitational-wave
emission mechanisms using several estimation techniques, including
two-dimensional numerical computation of quadrupole wave emission,
estimates of bar-mode strength, estimates of r-mode emission, and
estimates of waves from black hole ringing.  We also review the rate
at which the relevant collapses are believed to occur, which has a
major impact on their relevance as astrophysical sources.  Although
the latest supernova progenitor simulations produce cores rotating
much slower than those used in the past, we find that bar-mode and
r-mode instabilities from core-collapse supernovae remain among the
leading candidate sources for LIGO-II.  Accretion induced collapse
(AIC) of a white dwarf could produce gravitational-wave signals
similar to those from core-collapse.  In the models that we examine,
such collapses are not unstable to bar modes; we note that models
recently examined by Liu and Lindblom, which have slightly more
angular momentum, are certainly unstable to bar formation.  Because
AIC events are probably 1,000 times less common than core-collapse
supernovae, the typical AIC event will be much further away, and thus
the observed waves will be much weaker.  In the most optimistic
circumstances, we find it may be possible to detect gravitational
waves from the collapse of 300\,M\sun\/ Population III stars.

\end{abstract}


\section{Introduction}

We are entering an age where gravitational-wave (GW)
detectors will be sufficiently sensitive to observe a host
of astrophysical sources. This has led to a flurry of
activity among astrophysicists to estimate the
characteristics of sources of GW emission.  One source class
that has been the subject of much study is the collapse of
massive stars to form compact remnants (either neutron stars
or black holes).  Within this broad class, the collapse of
supernova progenitors (typically $\sim 15$\,M\sun\/ stars)
has received particularly strong scrutiny.  Most studies
have concentrated on calculating the GW emission during
collapse, or the emission from bar-like instabilities
shortly after bounce (see Rampp, M\"uller, \& Ruffert 1998
for a review).  However, the general class of stellar
collapse includes a rather wide variety of astrophysical
objects.  Such objects range from the collapse of a white
dwarf whose mass is pushed just beyond the Chandrasekhar
limit via accretion (accretion induced collapse, or AIC; see
Liu \& Lindblom 2001) to the collapse of very massive stars,
in excess of 260\,M\sun\/ (Fryer, Woosley, \& Heger 2001).

Although there is little doubt that stellar collapse produces GWs, it
is difficult to accurately estimate the characteristics of the signal
produced.  The amount of radiation depends sensitively upon the
rotation rate of the collapsing object.  The choice of pre-collapse
rotation can make a large difference in the resulting GW signal (Brown
2001), particularly in the case of bar-mode instability calculations
in core-collapse supernovae.

Simulations generally suffer from two deficiencies.  First, most
collapse simulations use simplified equations of state that do not
include the effects of neutrinos.  Although neutrinos may not be
important at early times (e.g., the simulations of Rampp et al. 1998),
at later times they strongly affect the material dynamics and cannot
be neglected.  In this paper we take advantage of the results from a
series of recent stellar-collapse simulations (Fryer et al. 1999a;
Fryer \& Heger 2000; Fryer et al. 2001) which followed
stellar collapse to late times with realistic equations of state and
neutrino physics.

Second, until recently no models of the progenitors of stellar
collapse have included the effects of rotation, so that there has
been little information available to constrain their initial spin.
Recently, however, Heger (1998) has developed a stellar evolution code
which includes a number of angular momentum transport processes and
can evolve massive stars to collapse.  He found that the rotation
speeds of collapsing supernova cores were much smaller than those used
in most calculations of GW emission from core-collapse (e.g. Rampp et
al. 1998).  In this paper, we re-investigate the GW emission from
stellar collapse using these latest progenitors in an effort to
calculate a GW signal which more closely reflects what is produced in
nature.

Our goal in this analysis is to make reasonable estimates
for the GW strength from stellar collapse, and in particular
to identify which scenarios lead to interesting GW sources
for detectors such as LIGO.  For each of the scenarios that
we consider (accretion induced collapse, core-collapse
supernovae, collapse of $M\gtrsim 300$\,M\sun\/ stars) we
examine the GW emission from bulk mass motions, from mass
currents, and from the ``ringing'' of a black hole (if one
is produced).  In this analysis we cannot precisely model GW
production---our axisymmetric code cannot follow
3-dimensional instabilities, nor the complicated behavior
for very long after collapse.  We thus make a number of
important assumptions and approximations to estimate GW
characteristics, sufficient to accomplish the goal of this
paper: to derive reasonable estimates for the frequency and
strain of GW output.

GWs from mass motion are computed in three ways.  First, we
numerically evaluate the quadrupole moment and its rate of change from
the axisymmetric simulations.  These waves come from polar-type
oscillations, and do not account for the instabilities caused by
rotation.  This axisymmetric estimate should strongly underestimate
the GW emission from a source.  Second, we assume the evolution of the
system may lead to a bar-mode instability, and calculate the GW
emission from the bar-mode that might develop.  There are indications
that the mass and angular momentum of some of our systems are in the
range where secular, and perhaps dynamical, bar-mode instabilities
develop.  We estimate the bar GW strength over a range of evolutionary
outcomes.  Finally, as a physically motivated upper limit to GW
production, we consider a fragmentation instability, wherein the
star's interior fragments into clumps.  We model these clumps as a
binary system.  This binary is a rather strong radiator, and yields a
reasonable upper bound on the GWs possible from a stellar source.  It
is not clear if such an instability would actually occur in a
realistic collapse, but it has not been ruled out (Fryer et al.
2001).  Indeed, we note that Van Putten (2001) has recently argued
that a similar instability may be a source of copious GWs during long
duration gamma ray bursts.

The r-mode instability produces GWs from mass currents.
We estimate the r-mode wave characteristics following the approach of
Ho \& Lai (2000).  This instability can be activated in a newly born, hot
neutron star.  It may also be reawakened when material falls back onto
the neutron star after it has cooled---the fallback material heats
and spins up the star.

In some evolutionary scenarios a black hole is formed.  The nascent
black hole is likely to be quite distorted from the quiescent Kerr
form, and remains so as material falls back onto it in the first
seconds or minutes of its life.  This distortion drives the hole to
radiate as it settles down to a Kerr state; the emitted waves are called
``ringing'' waves since they are qualitatively similar to the ringing
of a bell.  We estimate the strength and detectability of these waves
using some simplifying assumptions about the distribution of infalling
matter and the manner in which it distorts the hole.

In all cases we compare the expected wave strain from our sources with
the noise in the broad-band configuration of enhanced LIGO
interferometers (``LIGO-II'', see Gustafson et al. 1999).  The
comparison is based on a measure of the characteristic noise strain
which assumes good knowledge of the source's characteristics
(``matched filtering''; see Appendix A for further details).  It is
worth noting that other data analysis techniques are likely to be very
useful in searching for these waves (for example, the $f-\dot f$
technique of Van Putten \& Sarkar (2000) would probably do very well
at following the evolution of the black hole ringing frequency as its
mass and spin evolve due to accretion).  Since we are only interested
in a first broad discussion of these waves, we do not consider these
other data analysis techniques here.

LIGO is just one of several ground-based gravititational-wave
interferometers currently planned or under construction.  In Europe, a
French-Italian collaboration is building the 3 kilometer VIRGO
interferometer near Pisa, Italy (\cite{virgo} and references therein).
It will operate in the same time frame as LIGO, and is expected to
have very similar noise characteristics.  [One exception is that VIRGO
will use a very sophisticated seismic isolation system that promises
to move the low frequency noise ``wall'' from about 10 Hz (LIGO) to
roughly 3 or 4 Hz.]  A British-German collaboration is building the
600 meter GEO600 interferometer near Hannover, Germany (\cite{geo600}
and references therein).  One of GEO's major goals is to use advanced
interferometry techniques and technology from the beginning.  As well
as serving as a useful testbed for design ideas that will be used to
improve LIGO and VIRGO, this design compensates for GEO's shorter arms
and enables it to achieve sensitivity comparable to that of the
multi-kilometer instruments.  In Japan, the TAMA collaboration is
currently operating a 300 meter interferometer near Tokyo (\cite{tama}
and references therein).  This interferometer is being used as a
testbed for a future multi-kilometer instrument with cryogenically
cooled mirrors that they hope to build in the Kamioka mine
(\cite{kurodaetal}).  TAMA is of sufficient sensitivity to detect
gravitational-wave events within the Milky Way and nearby galaxies.
Finally, there are plans to build a LIGO-scale detector in western
Australia, near Perth (\cite{australia}).  This would be a
particularly valuable addition to the stable of detectors since the
Northern Hemisphere detectors lie very nearly within a common plane.
An Australian detector would be far outside of this plane, allowing it
to play an important role in determining the location of sources on
the sky.

In the frequency band of greatest interest to this analysis ($f\gtrsim
100$ Hz or so), the different interferometers have sensitivities that
are very similar to one another.  Our discussion, which focuses on
LIGO-II, carries over with little change to the other instruments.  At
low frequencies, VIRGO may have some advantage because of their
aggressive seismic isolation design; in particular, they may have a
better chance of detecting waves from the death of population III
stars (which are at low frequencies because of the cosmological
redshift).  In any case, we provide enough information for the
interested reader to compare our estimated wavestrains with the
sensitivity of detectors other than LIGO-II.

The remainder of this paper is organized as follows.  We review the
various mechanisms that lead to GW emission in \S 2, and then apply
them to collapse progenitors in \S 3 (AIC), \S 4 (core-collapse
supernovae), and \S 5 (collapse of very massive stars).  In each of
these sections we review current constraints on collapse rates and
discuss the detectability of the GWs. Background for the detectability
discussion is reviewed in Appendix A. In several places we discuss
sources at cosmological distances.  To convert between redshift and
luminosity distance (\cite{hogg}), we assume a flat universe with
$\Omega_m = 0.35$ and $\Omega_\Lambda = 0.65$, and with a Hubble
parameter $h_{100} = 0.65$. (Bahcall et al. 1999). A summary of the
results concludes the paper.

\section{Gravitational Wave Emission Mechanisms}

The collapse of massive stars involves a large amount of mass ($\sim
1$--$100\ M_\odot$), in a fairly compact region ($\sim
10^8$--$10^9\/\mbox{cm}$), moving at relativistic velocities ($v/c\sim
1/5$)---precisely the conditions needed for strong GW generation.  In
what follows we will explore a number of GW emission mechanisms,
including large-scale mass flows (quadrupole oscillations, bar-mode
and fragmentation instabilities), large-scale mass currents (r-mode
instability), and emission from the ringing of a newly-formed black
hole.  Each mechanism operates in a different regime, and the nature
of GW emission depends sensitively upon the evolutionary development
of the source.  We will consider each mechanism in turn, discussing
when each becomes important, and estimating the strength of the
resulting gravitational radiation.

We begin with a very brief review of GW theory; further detail and
references can be found in Thorne (1987).  The conventional
approach to calculating the GW emission of a given mass distribution
is via a multipole expansion of the perturbation $h_{\mu\nu}$ to a
background spacetime $g_{\mu\nu}^{\rm B}$.  The transverse-traceless
projection of this metric, evaluated in the radiation zone, is the
metric of the radiation field.  The lowest (quadrupole) order piece of
this field is (Thorne 1980)
\begin{equation}
h_{jk}^{\rm TT}=\left[{
{2\over d}{G\over c^4} {d^2\over dt^2}{\cal I}_{jk}(t-r)
+{8\over3d}{G\over c^5}\epsilon_{pq(j}{d^2\over dt^2}
{\cal S}_{k)p} (t-r)n_q}\right]^{\rm TT}.
\label{eq:thorne1}
\end{equation}
Here, ${{\cal I}_{jk}}$ and ${{\cal S}_{jk}}$ are the mass and current
quadrupole moments of the source, $d$ is the distance from the source
to the point of measurement, $\epsilon_{ijk}$ is the antisymmetric
tensor, and $n_q$ is the unit vector pointing in the propagation
direction.  Parentheses in the subscripts indicate symmetrization over
the enclosed indices, and the superscript {\rm TT} indicates that one
is to take the transverse-traceless projection; $G$ is Newton's
gravitational constant, and $c$ is the speed of light.

Most GW estimates are based on Eq.\ (\ref{eq:thorne1}).  When bulk
mass motions dominate the dynamics, the first term describes the
radiation.  For example, this term gives the well-known ``chirp''
associated with binary inspiral.  We will use it to model bar-mode and
fragmentation instabilities.  At least conceptually, this term also
applies to black hole ringing, provided one interprets ${\cal I}_{jk}$
as a moment of the spacetime rather than as a mass moment
(\cite{membrane}).  In practice, ringing waves are computed by finding
solutions to the wave equation for gravitational radiation
(\cite{teuk}) with appropriate boundary conditions (radiation purely
ingoing at the hole's event horizon, purely outgoing at infinity; see
Leaver 1985 for further discussion).  The second term in Eq.\
(\ref{eq:thorne1}) gives radiation from mass currents, and
is used to calculate GW emission due to the r-mode instability.

When the background spacetime is flat (or nearly so) the mass and
current moments have particularly simple forms.  For example, in
Cartesian coordinates the mass quadrupole is given by
\begin{equation}
{\cal I}_{jk} =
\int d^3x\,\rho\left[{x^jx^k-{1\over3}r^2\delta_{jk}}\right],
\end{equation}
where $\rho$ is the mass density, and $\delta_{jk}=1$ for $j=k$ and
$0$ otherwise.  The second term in the integrand ensures that the
resulting tensor is trace free.

Gravitational waves carry away energy and angular momentum from the source
(\cite{isaacson}).  The lowest order contribution to the power $P$
emitted in GWs is due to variations in the quadrupole moment:
\begin{equation}
P={dE\over dt}={1\over5}{c^5\over G}
\left\langle{\dddot{{\cal I}_{jk}}\dddot{{\cal
I}_{jk}}}\right\rangle,
\label{quad}
\end{equation}
where the dots refer to time derivatives.  For the purpose of
detectability estimates, it is usually more important to know the
dimensionless strain $h$ associated with a source.  This strain gives
the fractional change in distance between two separated masses as a GW
passes by; it is the quantity that is directly measured by GW
detectors.  The tensor field $h_{jk}^{\rm TT}$ encodes two
polarizations, $h_+$ and $h_\times$.  These polarizations can pulled
out of $h_{jk}^{\rm TT}$ with appropriate projection operators (see
{\cite{300years}} for expressions and further discussion).  The RMS
strain associated with a source is then given by $h =
\sqrt{\langle h_+\/^2 + h_\times\/^2 \rangle}$, where the angle
brackets denote both an average over several wavelengths and an
average over the sky.  This RMS strain will be used in all of our
analyses.

In general, the relation between strain and power scales as
\begin{equation}
P= {\pi^2 c^3\over G}f^2d^2h^2,
\end{equation}
where $f$ is the GW frequency and $d$ is the luminosity distance to
the source.  For a given strain, higher frequency waves radiate more
energy.  Because of detector noise, however, higher frequency waves
are not necessarily more detectable.

In the remainder of this section we discuss five different
approximation methods which are valid under varying conditions during
stellar collapse.  The first three (numerical quadrupole formalism,
bar mode formation, fragmentation instability) can be used to get a
handle on GW emission during the collapse itself, while the latter two
(r-modes and black hole ringing) occur after the formation of a
compact remnant.

\subsection{Numerical quadrupole formalism}

It is possible to directly apply Eq.~(\ref{quad}) to the results of a
numerical simulation.  By evaluating the quadrupole moment on multiple
time slices, one can compute the time derivatives numerically, and
thereby determine the GW emission.  Computing time derivatives across
many slices, however, can generate an unacceptable amount of numerical
noise.  An alternate approach, based upon work by Blanchet et
al. (1990), and used extensively by Zwerger \& M\"uller (1997),
rewrites the time derivatives as spatial derivatives, thereby avoiding
the need to consider multiple time slices when calculating
instantaneous power emission.  Both the velocities of the particles
and the Newtonian potential are known on a given slice.  The gradient
of the potential yields the forces acting on the particles, and,
coupled with the velocity information, determines what the next slice
will look like.  Therefore, by utilizing the gradients directly, it is
possible to calculate the GW emission while restricting oneself to a
single numerical slice.

The expression we use to do this in the axisymmetric case comes from
Zwerger \& M\"uller (1997). The only non-vanishing quadrupole wave
amplitude component, $A_{20}$, is given by
\begin{eqnarray}
A_{20}(t)&=&{G\over c^4}{16\pi^{3/2}\over\sqrt{15}}
\int_{-1}^1\int_0^\infty\,dr\,dz\,r^2\rho
\Big[{v_r^2(3z^2-1)+v_\theta^2(2-3z^2)}\nonumber\\
&&{\quad{}-v_\phi^2
-6v_r v_\theta z\sqrt{1-z^2}
- r{\partial\Phi\over\partial r}(3z^2-1)
+ 3 {\partial\Phi\over\partial\theta}z \sqrt{1-z^2}}\Big]\;.
\label{zm_eq}
\end{eqnarray}
Spherical coordinates are used here, with $z = \cos\theta$;
$\Phi$ is the Newtonian gravitational potential. The
non-zero wave strain component is then given by
$h_+=\sqrt{(15/64\pi)}\sin^2\!\theta\,A_{20}/d$, where $d$ is
the luminosity distance to the source.

Although our code is axially symmetric (such that all the
$\partial_\phi$ terms vanish), Eq.\ (\ref{zm_eq}) still yields
non-zero GW emission due to polar-type oscillations.  These modes
become important when there is large {\em aspherical} mass infall (or
ejection).  However, we would strongly underestimate
GW emission if we restricted ourselves to these modes---emission
should be much stronger in cases where the tangential motion of
particles dominates the radial motion.  We now discuss a number of
non-axisymmetric modes which our code is unable to reproduce 
directly.

\subsection{Bar modes}

A bar mode instability is one of the more promising mechanisms by
which a significant fraction of the collapsing system's energy can be
emitted in GWs.  Bar mode instabilities occur in objects whose
rotational kinetic energy exceeds some fraction of their potential
energy, with the ratio generally written as $\beta \equiv T/|W|$.
Standard lore (Chandrasekhar 1969) states that an object is unstable
on a secular time scale if $\beta\gtrsim 0.14$, and is dynamically
unstable if $\beta\gtrsim 0.27$.  [We note, though, that recent work
suggests that if the collapse hangs due to centrifugal forces
producing a density profile which is not centrally peaked, dynamical
instabilities can occur at a much lower value of $\beta$ (Centrella et
al. 2000).]

In most core-collapse simulations done with up-to-date progenitor
models, it is found that the mass distribution does not become
centrifugally hung up.  This suggests that instability to bar mode
formation might be unlikely: in these simulations, the density is
centrally concentrated, and thus the low critical values of Centrella
et al. (2000) do not apply.  High values of $\beta$ are necessary to
induce bar-mode instabilities.  Fortunately, the rotational energy can
be very high in these models (Fig. 1), and it remains likely that
bar-mode instabilities will occur.

Heartened by these results, we review here expressions describing bar
mode GW emission.  Consider a bar of mass $m$ and length $2r$,
rotating with angular frequency $\omega$. The GW energy radiated is
given, in the quadrupole approximation, by
\begin{equation}
P_{\rm bar}={32\over45}{G\over c^5}m^2r^4\omega^6.
\end{equation}
A detector at a distance $d$ from the source would measure an rms
strain
\begin{equation}
h_{\rm bar}=\sqrt{32\over45}{G\over c^4} {m r^2 \omega^2\over d}.
\end{equation}
Note that, due to symmetry, the frequency of the emitted GWs is twice
the bar's rotation frequency.

\subsection{Fragmentation instability}

To set a physically motivated upper limit to the GW emission that
might be produced in stellar collapse, we imagine that the collapse
material fragments into clumps, which then orbit for some number of
cycles as the collapse proceeds.  For concreteness we consider the
material fragmenting into a binary system, though it could very well
fragment into more objects.  We note that collapse simulations give
some indication that this kind of instability may be present
(\cite{FWH}).  Also, Van Putten has recently argued that a
fragmentation-type instability in collapsar powered gamma-ray bursts
may drive very strong GW emission (\cite{VP2001}).

For two bodies, each of mass $m$, in circular orbit about one another
at a frequency $\omega$ and with separation $2r$, the power and mean
strain are given by:
\begin{eqnarray}
P_{\rm bin} &=& {128\over5}{G\over c^5}{m^2 r^4\omega^6}\\
h_{\rm bin} &=&
\sqrt{128\over5}{G\over c^4}{m r^2\omega^2\over d}.
\end{eqnarray}
These results make no assumption about orbital frequency, and so apply
to, for example, pressure supported as well as Keplerian orbits. For
Keplerian orbits we have $4\omega^2 r^3=Gm$, and the above expressions become
\begin{eqnarray}
P_{\rm bin} &=& {2\over5}{G^4\over c^5}{m^5\over r^5}\\
h_{\rm bin} &=&
\sqrt{8\over5}{G^2\over c^4}{m^2\over r\,d}.
\end{eqnarray}
Note that if the ``horizons'' of the two bodies touch ($r=2m\,G/c^2$),
then the power radiated reaches a maximum of $P = c^5/80G \sim
10^{57}\mbox{ ergs s}^{-1}$, independent of the mass of the system.
The length of time such power emission is sustained, however, scales
with the total mass (thus supermassive black hole binaries radiate
more than microscopic ones).

\subsection{R-modes}

If the collapse forms a neutron star, an r-mode instability can arise
at late times, once the neutron star has cooled.  This instability has
attracted much attention in the past few years, fueled by the
discovery by Andersson (1998) and Friedman \& Morsink (1998) that
gravitational radiation drives r-mode instabilities in rotating
neutron stars [see Lindblom (2001) for a review].  Unfortunately, an
accurate calculation of GWs from the r-mode instability requires an
understanding of the growth and maximum limit of the r-mode amplitude
which, in turn, requires an understanding of the viscous terms that
act to damp the modes (e.g., shear and bulk viscosities in the neutron
star fluid, shear viscosities caused by crusts, magnetic viscosities,
etc.).  For instance, Rezzolla, Lamb, and Shapiro (2000) found that
poloidal fields as low as $10^{10}$\,G could damp out r-modes.  These
viscous terms depend upon the neutron star structure and equation of
state.  On top of these uncertainties, the GW signal depends upon the
formation and evolution of young neutron stars (especially during the
first 1000\,s), and hence is subject to the various uncertainties
associated with those processes.

For our study of GWs from r-modes, we use the simplified equations
derived by Ho \& Lai (2000), which consider only the dominant $l=m=2$
mode with the initial neutron star structure from Owen et al. (1998):
$M_{\rm NS}$=1.4\,M\sun\/, $R_{\rm NS}$=12.53\,km (which follows from a
polytropic equation of state with index $\Gamma=2$).  The spin
frequency ($\nu_{\rm S}$) evolution of the neutron star from Ho \& Lai
(2000) is:
\begin{equation}
\frac{d \nu_{\rm S}}{dt} = - 2 Q \frac{\nu_{\rm S} 
\alpha^2}{\tau_{\rm V}}
\end{equation}
where the viscous timescale ($\tau_{\rm V}$) is given by:
\begin{equation}
\frac{1}{\tau_{V}} = \frac{T_9^{-2}}{2.52 \times 10^8}+ 
\frac{T_9^6}{1.26\times10^9}
\end{equation} 
with $Q=0.094$ and the neutron star temperature $T_9=T/10^9$\,K.  
For most of our calculations we assume that the temperature evolves
with time under the prescription of Owen et al. (1998): $T_9=(t/1
\mbox{ yr})^{-1/6}$.  The r-mode amplitude $\alpha$ is driven by gravitational
radiation and damped by viscous forces:
\begin{equation}
\frac{d \alpha}{dt} = - \frac{\alpha}{\tau_{\rm GR}} 
- \frac{\alpha}{\tau_{\rm V}} (1-\alpha^2 Q)\;.
\end{equation}
The gravitational radiation timescale ($\tau_{\rm GR}$) is given by:
\begin{equation}
\frac{1}{\tau_{\rm GR}} = - \frac{1}{18.9\,{\rm sec}}
\left ( \frac{\nu_{\rm S}}{1\,{\rm kHz}} \right )^6.
\end{equation}
With this prescription the r-mode amplitude would grow to very large
values unless we set some critical amplitude.  As in Ho \& Lai (2000),
if the mode amplitude rises above 1, we hold it constant and the spin
evolution becomes:
\begin{equation}
\frac{d \nu_{\rm S}}{dt} = - 2 \frac{\nu_{\rm S} 
\alpha^2}{\tau_{\rm V}} \frac{\alpha^2 Q}{1-\alpha^2 Q}.
\end{equation}
Ho \& Lai included the effects of magnetic breaking (via radiation
from a dipole magnetic field) in the newly formed pulsar, and we
discuss these effects in \S\S 3 and 4.

We also examine the effect of fallback accretion onto the collapsed
remnant.  In core-collapse supernovae it is likely that $\gtrsim
0.1$\,M\sun\/ of material will fall back onto the newly formed neutron
star $\sim20-2000$\,s after the explosion.  The angular momentum of
this fallback material leads to the formation of an accretion disk which
spins up the neutron star as it accretes:
\begin{equation}
\frac{d \nu_{\rm S}}{dt}_{\rm Fallback}=
\frac{d {J_{\rm Fallback}}/{dt}-2 \pi \nu_{\rm S} dI/dt}{2 \pi I}.
\end{equation}
We assume that the angular momentum per unit mass accreted onto the
star is equal to that of a Keplerian orbit at the neutron star
surface:
\begin{equation}
\frac{d J_{\rm Fallback}}{dt} = \frac{dM_{\rm
Fallback}}{dt} \sqrt{G M_{\rm NS} R_{\rm NS}}.  
\end{equation}
We also assume that the change in the moment of inertia ($I$) is
limited to the change in mass (the radius remains constant):
\begin{equation}
dI/dt= \frac{2 \pi \nu_{\rm S} I}{M_{\rm NS}}  \frac{dM_{\rm
Fallback}}{dt}.
\end{equation}
This assumption is adequate to get the qualitative flavor of the 
accretion effects on GW emission from r-modes.

Typical fallback accretion rates for supernovae which produce neutron
stars peak in the range $3 \times 10^{-4}$--$3 \times
10^{-3}$\,M\sun\,s$^{-1}$ (Fryer, Colgate, \& Pinto 1999b).  Such high
rates quickly smother any magnetic field, heating up the neutron star.
The neutron star temperature is set by the temperature at the surface
of the neutron star (Fryer et al. 1999b):
\begin{equation}
T_9=2.16\times 10^{3} S^{-1} \frac{10^6 {\rm cm}}{R_{\rm NS}},
\end{equation}
where $S$ is the entropy of the infalling material.  For a
$\Gamma=4/3$ polytrope, this is given by (Fryer et al. 1999b):
\begin{equation}
S=11.8
\left ( \frac{M_{\rm NS}}{M_{\scriptscriptstyle\odot}} \right )^{7/8} 
\left ( \frac{M_{\scriptscriptstyle\odot}\mbox{ s}^{-1}}{dM/dt} \right )^{1/4} 
\left ( \frac{10^6\ {\rm cm}}{R_{\rm NS}} \right )^{3/8}
k_{\rm B}{\rm \,  per\, nucleon}.
\end{equation}

Combining the equations for the evolution of the neutron star spin
and the r-mode amplitude with Ho \& Lai's (2000) definition for the
average wave amplitude $h(t)$ yields
\begin{equation}
h(t) = 1.8\times10^{-24}\alpha \left({\nu_{\rm S}\over{\rm 1 kHz}}\right)
\left({20\,{\rm Mpc}\over d}\right)\;.
\end{equation}
We use this formula to estimate r-mode GW emission in newly collapsed
stars.

\subsection{Black Hole Ringing}

If the stellar collapse forms a black hole instead of a neutron star,
a different mechanism produces GWs.  The properties of the black hole
during collapse rapidly change as material from the star falls onto
it, increasing its mass and possibly its spin.  The infalling matter
also perturbs the hole's geometry, distorting it from the Kerr
solution.  This distortion causes the hole to ``ring'' in distinct
harmonics as gravitational radiation carries away the perturbation and
the hole settles into a quiescent, stationary Kerr state.

An accurate calculation of this ringing would require a code that
calculates the perturbation spectrum given a mass inflow.  Here we
take a much simpler approach, approximating the spectrum as a
stochastic superposition of Kerr quasi-normal modes arising from
repeated ``thumping'' by the matter flow onto the nascent black hole.
Although this analysis is not adequate to rigorously detail the
characteristics of the ringing waves emitted during massive star
collapse, it should be adequate to estimate the waves' strength and
detectability.  We hope to motivate more careful analyses that use
inflow codes to compute the emitted waves (an early example of which
is described in Papadopoulos \& Font 2001).

A black hole distortion can be decomposed into spheroidal
modes with spherical-harmonic-like indices $l$ and $m$.  The
quadrupole modes ($l = 2$) presumably dominate, while the
dominant $m$ value depends upon the matter flow.  The $m =
\pm 2$ modes are bar-like, co-rotating ($+$) and
counter-rotating ($-$) with the hole's spin; the $m =
\pm 1$ mode represents a shift of the black hole's position; and the $m = 0$
mode is an axisymmetric distortion.  For simplicity, we assume that the
hole rings entirely in $m = 0$ and $m = 2$ modes.  We exclude $m =\pm 1$
because it is difficult to estimate their importance without detailed
calculations of the ringing spectrum, and we exclude $m = -2$ because it
is likely to be strongly suppressed [it is counter to the spacetime's
rotation, and has an extremely short lifetime (see Chandrasekhar 1983,
Fig.\ 45)].  We will vary the fraction of energy radiated in $m = 0$ and
$m = 2$ waves to see how the detectability of these events varies
with the nature of the hole's distortion.

Each quasi-normal mode has a unique frequency $f_{lm}$ and damping
time $\tau_{lm}$, depending only on the hole's mass and spin
({\cite{leaver}}).  Useful fits to Leaver's numerical data (see Leaver
1985, Tables 2 and 3; also Echeverria 1989), accurate to $\sim 10\%$,
are
\begin{eqnarray}
f_0 &\simeq& 700\,{\rm Hz}\left(20 M_\odot\over M\right)\left[1 -
0.13(1 - a/M)^{6/10}\right]\;,\nonumber\\
Q_0 & \simeq & 3 - (1 - a/M)^{4/10}\;;
\label{eq:2_0_info}\\
f_2 &\simeq& 1600\,{\rm Hz}\left(20 M_\odot\over M\right)\left[1 -
0.63(1 - a/M)^{3/10}\right]\;,\nonumber\\
Q_2 & \simeq & 2(1 - a/M)^{-9/20}\;.
\label{eq:2_2_info}
\end{eqnarray}
The quality factor $Q \equiv \pi f \tau$.  We have suppressed the $l$
subscript since it is 2 in all cases.

The amplitude of the ringdown waves, and the energy that they carry,
depends upon the manner and extent to which the hole is distorted.  A
useful starting point is the ``DRPP'' result ({\cite{DRPP}}): the
total energy carried off by GWs when a mass $\mu$ falls radially onto
a Schwarzschild black hole of mass $M$ is
\begin{equation}
\Delta E_{\rm DRPP} = 0.01 \mu c^2(\mu/M)\;.
\label{eq:DRPP}
\end{equation}
For a rotating hole, this formula underestimates the energy radiated
for matter falling down the poles by a spin-dependent factor $\lesssim
50\%$ ({\cite{sasaknak}}).  Also, it only applies to an $m = 0$
perturbation---the energy radiated from a non-axisymmetric
perturbation could be significantly larger.  We will assume that Eq.\
(\ref{eq:DRPP}) correctly describes the scaling of $\Delta E$ with
$\mu$, but will allow the size of $\Delta E$ to vary with a parameter
$\varepsilon$:
\begin{equation}
\Delta E = \varepsilon \mu c^2(\mu/M)\;.
\label{eq:deltaE}
\end{equation}

From the energy emitted, $\Delta E$, we estimate the GW amplitude
associated with a single clump falling into the hole.  The energy flux
carried by GWs is ({\cite{isaacson}}):
\begin{equation}
{dE\over dA dt} = {c^3\over16\pi G}\overline{\left[\left({\partial
h_+\over\partial t}\right)^2 + \left({\partial h_\times\over\partial
t}\right)^2\right]}\;.
\label{eq:isaacsonflux}
\end{equation}
(The overbar indicates that this expression must be averaged over
several cycles or wavelengths.)  Quasi-normal ringing waves are damped
sinusoids, so the waveform for a combination of $m = 0$ and $m = 2$
waves can be written
\begin{eqnarray}
h_+(t) &=& \left[{\cal A}_0 S_{20}(\theta,\phi;a)e^{-t/\tau_0}
\cos(2 \pi f_0 t + \varphi_0)
+ {\cal A}_2 S_{22}(\theta,\phi;a)e^{-t/\tau_2}
\cos(2 \pi f_2 t + \varphi_2)\right]/D,\nonumber\\
h_\times(t) &=& \left[{\cal A}_0 S_{20}(\theta,\phi;a)e^{-t/\tau_0}
\sin(2 \pi f_0 t + \varphi_0)
+ {\cal A}_2 S_{22}(\theta,\phi;a)e^{-t/\tau_2}
\sin(2 \pi f_2 t + \varphi_2) \right]/D,\nonumber\\
\label{eq:waveforms}
\end{eqnarray}
where $D$ is the luminosity distance from the distorted hole, $\theta$
and $\phi$ are angles on the sky, the functions $S_{20}(\theta,\phi;a)$
and $S_{22}(\theta,\phi;a)$ are spheroidal harmonics [reducing to
spherical harmonics when the black hole is not spinning ($a = 0$)], and
($\varphi_0$, $\varphi_2$) are phase offsets, related to the initial
perturbation of the hole.

Plugging Eq.\ (\ref{eq:waveforms}) into Eq.\ (\ref{eq:isaacsonflux})
and integrating over a large sphere gives
\begin{equation}
{dE\over dt} = {c^3\over G}{\pi\over 4}
\left[{\cal A}_0^2 f_0^2 e^{-2t/\tau_0} +
{\cal A}_2^2 f_2^2 e^{-2t/\tau_2}\right]\;.
\label{eq:dEringdt}
\end{equation}
We have used the fact that the spheroidal harmonics are orthonormal
functions on the sphere, and we have approximated $2 \pi f \gg 1/\tau$
(which introduces errors of $\sim 10\%$).  Integrating over time and
using the definition of $Q$ yields
\begin{eqnarray}
\Delta E &=& {c^3\over8G}
\left[Q_0 f_0 {\cal A}_0^2 + Q_2 f_2 {\cal A}_2^2\right]\;,
\nonumber\\
&\equiv& \Delta E_0 + \Delta E_2\;.
\label{eq:deltaE_twomode}
\end{eqnarray}
We equate this $\Delta E$ to that given by Eq.~(\ref{eq:deltaE}).  Since
we do not know how to apportion this energy among the $m = 0$ and $m =
2$ modes, we split it up with a parameter $\alpha_{\rm ring}$:
\begin{equation}
\Delta E_0 = \alpha_{\rm ring} \Delta E\;,\qquad
\Delta E_2 = (1 - \alpha_{\rm ring})\Delta E\;.
\label{eq:variation}
\end{equation}
In principle, $\alpha_{\rm ring}$ could be time dependent---for
instance, it will increase if the mass inflow becomes axisymmetric later
in the collapse.  We ignore this possibility here, and take $\alpha_{\rm
ring}$ to be constant.  We finally obtain
\begin{eqnarray}
{\cal A}_0 &=& \sqrt{8G\alpha_{\rm ring}\Delta E\over Q_0 f_0 c^3} =
\sqrt{8\varepsilon G\alpha_{\rm ring}\mu^2\over Q_0 c M f_0}\;,
\nonumber\\
{\cal A}_2 &=& \sqrt{8G(1 - \alpha_{\rm ring})\Delta E\over Q_2 f_2 c^3} =
\sqrt{8\varepsilon G(1 - \alpha_{\rm ring})\mu^2\over Q_2 c M f_2}\;.
\label{eq:amp1}
\end{eqnarray}

To set an upper limit on the strength of the emitted waves,
we assume that the accretion flow is extremely clumpy: the
hole gets ``thumped'' by a clump of mass $\mu = {\dot
m}T_{\rm thump}$ every $T_{\rm thump}$.  The thump time
$T_{\rm thump}$ will be treated as a variational parameter:
for example, when it is very large, the ringdown signal
consists of a small number of very large thumps.  The
amplitudes for a single thump are
\begin{eqnarray}
{\cal A}_0 &=& \sqrt{8\varepsilon G\over c} {{\dot m}T_{\rm
thump}\over M^{1/2}}\left({\alpha_{\rm ring}\over Q_0}\right)^{1/2}\;,
\nonumber\\
{\cal A}_2 &=& \sqrt{8\varepsilon G\over c} {{\dot m}T_{\rm
thump}\over M^{1/2}}\left({1 - \alpha_{\rm ring}\over Q_2}\right)^{1/2}\;.
\label{eq:amp2}
\end{eqnarray}

We define the time index $k$ via $t_k = k T_{\rm thump}$.  By
stringing together a sequence of thumps and averaging over sky
position, we finally arrive at the following expression for the
gravitational waveform:
\begin{eqnarray}
h(t) \equiv h_+(t) - i h_\times(t) &=&
{\sqrt{2\varepsilon G/\pi c}\over D}
\sum_{k = 0}^{k_{\rm max}} {\dot m(t_k)T_{\rm thump}\over
M(t_k)^{1/2}}\times\nonumber\\
& &\left[\left[{\alpha_{\rm ring}\over Q_0(t_k)}\right]^{1/2}
e^{-(t - t_k)/\tau_0(t_k)}e^{-2 \pi i f_0(t_k)(t - t_k)}
e^{-i\varphi_0(t_k)}\right.\nonumber\\
&+&\left.\left[{1 - \alpha_{\rm ring}\over Q_2(t_k)}\right]^{1/2}
e^{-(t - t_k)/\tau_2(t_k)}e^{-2 \pi i f_2(t_k)(t - t_k)}
e^{-i\varphi_2(t_k)}\right]\;.
\nonumber\\
\label{eq:thump_waves}
\end{eqnarray}
All functions which evolve with time are written explicitly as functions
of $t_k$.  Note in particular the phases $\varphi_0(t_k)$ and
$\varphi_2(t_k)$: because they depend upon the distortion state of the
hole as each clump arrives and further distorts the horizon, they will
be random for all practical purposes.  Thus the ringing waves
emitted from the mass inflow will be stochastic.  The index $k_{\rm
max}$ describes the time at which the accretion flow ends and the hole
stops ringing.

Equation (\ref{eq:thump_waves}) will be coupled with descriptions of
the mass flow in later sections of this paper to estimate the
detectability of these waves.

\vspace{0.5cm}

Having finished our discussion of relevant GW emission mechanisms, in
the next three sections we turn to a number of possible astrophysically
relevant GW sources.

\section{Accretion Induced Collapse}

When a white dwarf's mass exceeds the Chandrasekhar limit\footnote{By
Chandrasekhar limit, we mean the maximum mass of a stable white dwarf.
Note that this depends upon composition and angular momentum.}, it
begins to collapse.  As it contracts, its temperature increases
adiabatically.  Neutrino cooling (via Urca processes) limits the rise
in temperature.  If neutrino cooling does not reduce the adiabatic
heating significantly, the collapsing white dwarf will reach
temperatures hot enough to ignite nuclear burning.  The entire white
dwarf explodes in a thermonuclear explosion known as a Type Ia
supernova.  If, on the other hand, cooling initially prevents nuclear
ignition, the white dwarf will collapse more and more quickly as
electrons capture onto protons, and the white dwarf will ultimately
form a neutron star.

This ``Accretion-Induced Collapse'' (AIC) of a white dwarf is very
similar to core-collapse supernovae.  The collapse of white dwarfs has
been studied in some detail over the past few decades (Hillebrandt,
Wolff, \& Nomoto 1984, Woosley \& Baron 1992, Fryer et al. 1999a), and
we have some understanding of the collapse process and the resultant
explosion.  Since the white dwarf is pushed over the Chandrasekhar
mass limit through disk accretion, it is likely that the collapsing
white dwarf will rotate at a significant rate, allowing for the
possibility of a variety of instabilities, and the concomitant
emission of GWs.  The analysis in this paper relies upon the rotating
simulation (model 3) of Fryer et al. (1999a).

\subsection{Formation Rate and Angular Momentum}

Calculating the formation rate of AICs from first principles is
fraught with difficulties, ranging from understanding
binary star evolution to uncertainties in the accretion process
itself.  We have already mentioned one such uncertainty: Does the star
ignite in a thermonuclear explosion or collapse to form an AIC?  At
present we have only a rudimentary understanding of the conditions
necessary for a white dwarf to gain matter during the accretion
process (as opposed to losing matter via a series of nova explosions).
Fortunately, due to the importance of Type Ia supernovae as
cosmological candles, there has been a lot of activity studying the
progenitor evolution and mass accretion necessary to produce
Chandrasekhar-massed white dwarfs (see Nomoto, Iwamoto, Kishimoto
1997; Branch 1998; or Livio 2000 for reviews).  Unfortunately, whether
or not the current mass-transfer progenitors actually lead to an
increase in the white dwarf mass to the Chandrasekhar limit is still a
matter of hot debate (contrast Kato \& Hachisu 1999 with Cassisi, Iben
\& Tornamb\'e 1998).  All of the uncertainties in binary evolution,
accretion, etc.~make it difficult at this time to directly calculate
the formation rate of Chandrasekhar-mass white dwarfs.

We therefore rely upon indirect methods to place constraints on the AIC
rate.  One way is to derive a rate of thermonuclear explosions from
Chandrasekhar-massed white dwarfs and calculate the {\it relative}
fraction of Chandrasekhar-massed white dwarfs which produce AICs.
Calculating relative rates removes some portion of the uncertainties,
and may be easier to estimate theoretically.  It is becoming
increasingly evident that the thermonuclear explosion of a
Chandrasekhar-massed white dwarf is the mechanism which produces Type Ia
supernovae (Nomoto, Iwamoto, \& Kishimoto 1997; Branch 1998; or Livio
2000).  From observations of Type Ia supernovae, we infer a minimum rate
at which Chandrasekhar-massed white dwarfs are produced in the Galaxy:
$4 \pm 1 \times 10^{-3}\,{\rm yr^{-1}}$ (Cappellaro et al. 1997).  Some
fraction of Chandrasekhar-massed white dwarfs will collapse to form a
neutron star.  Which fate befalls the white dwarf depends sensitively
upon the initial mass of the white dwarf, its chemical composition, and
the rate at which it accretes matter (see Nomoto \& Kondo 1991 for
review).  Although calculating this fraction may be easier than
calculating a rate directly, at present we are unable to make
accurate estimates.

By modeling the collapse itself, however, we can place constraints on
the AIC rate.  During the collapse the white dwarf ejects the outer
$\sim$0.1\,M\sun\/ of its envelope, some of which had become very
neutron rich due to electron capture.  As this material is ejected, it
``pollutes'' the Galaxy with extremely rare, neutron-rich isotopes.
By measuring the total amount of these isotopes in the Galaxy, and
assuming these isotopes are formed solely in AICs, we can place an
upper limit on the rate of AICs in the Galaxy at about $10^{-5}$ per
year (Fryer et al. 1999a).

What about the angular momentum distribution of these collapsing 
stars?  The white dwarf is pushed above the Chandrasekhar mass 
either by accretion or through the merger of two white dwarfs.  
In either case the process by which the white dwarf gains mass 
also causes the white dwarf to gain a considerable amount of 
angular momentum.  One can roughly estimate the rate of angular 
momentum gain by assuming all of the angular momentum at 
the inner edge of the accretion disk is added to the white dwarf:
\begin{equation}
\dot{J}=\dot{m} \sqrt{G M_{\rm WD} R_{\rm disk}},
\end{equation}
where $\dot{m}$ is the accretion rate, $G$ is the gravitational
constant, $M_{\rm WD}$ is the mass of the white dwarf, and $R_{\rm
disk}$ is the inner radius of the accretion disk.  A 2500\,km,
1.35M\sun\/ white dwarf (with $R_{\rm disk}$ set to the white dwarf
radius) accreting 0.05\,M\sun\, gains $2\times 10^{49} {\rm\,g\,cm^2
\,s^{-1}}$ in angular momentum.  Given the small moment of inertia of
white dwarfs (typically, $I/M_{\rm WD}R^2_{\rm WD} < 0.1$), this small
amount of accretion can cause an initially non-rotating neutron star
to achieve rotation rates greater than 1\,rad\,s$^{-1}$, or periods
less than 6\,s.  It should be noted, though, that rotation periods for
cataclysmic variables typically range from 200--1200\,s (Liebert 1980)
[although periods as low as $\sim 30$\,s exist (King \& Lasota 1991)].
Most white dwarfs lose mass during accretion due to novae, and this
may limit the amount of angular momentum accreted (e.g. King, Wynn, \&
Regev 1991).  It is possible that those white dwarfs which actually
gain mass up to the Chandrasekhar mass may have much faster spin
periods than the observed sample.  Even so, the actual spin period
before collapse depends upon a variety of uncertainties: white dwarf
radius, accretion rate, magnetic field strength, etc. (e.g. Narayan \&
Popham 1989).  For the purposes of our analysis, we assume that the
white dwarf collapsed with $J=10^{49} {\rm\, g\, cm^2\, s^{-1}}$
(Fig. 2).  Liu \& Lindblom (2001) have recently studied white dwarfs
with more angular momentum ($J=3$--$4\times10^{49} {\rm\, g\, cm^2\,
s^{-1}}$), and we will compare our results with theirs below.

\subsection{Gravitational Waves}

\subsubsection{GW from Collapse or Explosion}

As we discussed in \S 2, bar-mode instabilities are driven when the
ratio of rotational kinetic energy to potential energy ($\beta$)
exceeds some critical value.  Toroidal density distributions (where
the density peaks some distance away from the center of the object)
are more susceptible to bar-mode instabilities, and may develop these
instabilities for $\beta \gtrsim 0.10-0.12$ (Tohline \& Hachisu 1990;
Centrella et al. 2000).  The collapse of a rotating white dwarf does
produce a density distribution peaked away from the center (Fig. 3),
and the critical $\beta$ may be as low as Tohline \& Hachisu (1990)
predict.  However, for our choice of initial spin, $\beta$ is less than
0.06.  For AICs, the boundary between proto-neutron star and ejected
material is generally very sharp, but in core-collapse supernovae it
is sometimes difficult to define the edge of the proto-neutron star.
Hence we have calculated $\beta$ as a function of enclosed mass, where
$\beta$ corresponds to the total rotational energy and potential
energy in that enclosed mass.  Note that no matter where we define the
edge of the neutron star, $\beta < 0.06$.  That said, we should bear
in mind that the higher angular momentum white dwarfs of Liu \&
Lindblom (2001) are almost certainly unstable.

Polar-type oscillations estimated from the quadrupole formula (\S 2.1)
predict a peak strain of $5.9 \times 10^{-24}$ at 100\,Mpc and GW
energies of $3\times10^{45}$\,ergs (see Table 1).  These waves are
emitted at a frequency of about 50 Hz.  The RMS noise strain of
LIGO-II broad-band interferometers (cf.\ Appendix A) is about $6\times
10^{-23}$ near 50 Hz.  The strength of these waves is an encouraging
sign that, {\it if}\/ stronger instabilities such as bar mode formation
were to become active, they would likely be of great observational
importance (though these polar observations are not themselves
observationally interesting).

\subsubsection{GW from Remnants}

For the first 10--20\,s after collapse and explosion from the AIC, the
proto-neutron star remnant remains electron rich.  Electron neutrinos
(created via electron capture onto protons) deleptonize the neutron
star, but since the degeneracy energy of electrons is less than the
energy carried away by the neutrinos (that is, the energy released
through electron capture is more than the energy carried away by the
neutrinos produced in electron capture), the neutron star initially
heats up (Keil \& Janka 1995).  It can take up to 50\,s before the
temperature falls below $10^{10}$\,K.  By this time the proto-neutron
star has contracted to nearly its final radius.  Assuming a
1.4\,M\sun, 12.53\,km neutron star ($\Gamma=2$ polytrope), the moment
of inertia is only $1.1 \times 10^{45}$\,g\,cm$^2$.  As long as the
total angular momentum is above $7\times10^{48}{\rm\ g \, cm^2 \,
s^{-1}}$, the period will be below 1\,ms, and the neutron star remnant
will be an ideal candidate for strong r-mode emission.  However,
r-modes are highly damped until the temperature (and hence the bulk
viscosity) decreases below $10^{10}$\,K.  In our analysis, we assume
the neutron star has contracted after 10\,s and using $T_9=(t/1 {\rm\,
yr})^{-1/6}$, $T_9=12$ at this time.  As the proto-neutron star cools,
the r-mode amplitude grows and converts much of the rotational energy
into GW emission (Fig. 4; solid line).  To compare with Ho \& Lai
(2000), we have assumed a spin frequency set to 890Hz. If instead the
neutron star is initially much hotter, but cools faster ($T_9=(t/1
{\rm\,yr})^{-1/3}$), the GW signal occurs later, but the strength and
rough structure is nearly the same (Fig. 4; dot-dashed line).  The
total energy emitted in GWs exceeds $10^{52}$\,ergs with a maximum
power of over $10^{50}$\,ergs\,s$^{-1}$ (Table 1).

AIC r-mode waves are compared to LIGO's mean noise in the lower track
of Figure {\ref{fig:rmode}} (see Appendix A for discussion of how this
track is calculated, and Owen et al. (1998) for further discussion).
The track illustrates the strength of the waves if it were possible to
coherently integrate the signal's phase evolution over the course of 1
year.  The wave track is below the mean noise everywhere on the plot,
indicating that these waves are not detectable.  This is not too
surprising, since one typically discusses r-mode strength for sources
no more than about 20 Mpc away --- 100 Mpc is just too far for the
source to be detectable.

The above calculations assume that the neutron star has also acquired
a $10^{12}$\,G dipole magnetic field and is emitting as a pulsar 
(however, we neglect any damping effects caused by magnetic fields).  In
the first year, a total of $2\times10^{47}$\,ergs is lost through
pulsar emission, most of which will power the accelerated ejection of
the exploding material.  This energy is less than 0.1\% of the
explosion energy, and only 0.001\% of the total rotation energy.
Although this energy will not make a difference in the supernova light
curve at early times, the pulsar luminosity remains high long after
the supernova and GW emission has died away, and will easily be
visible after the ejecta has cleared.  A $10^{13}$\,G magnetic field
pulsar is similar ($E_{\rm Pulsar, 1yr} = 2\times10^{49}\,\mbox{ergs}
= 2\%\, E_{\rm explosion} = 0.1\%\, E_{\rm GW}$).  At $10^{14}$\,G,
the pulsar mechanism will produce observable effects in the supernova
explosion at peak, even though less than 10\% of the rotational energy
is going into pulsar emission.  If AICs do produce such highly
magnetized, rapidly rotating stars, we should observe the pulsar
outbursts.\footnote{We assume that the dipole magnetic field mechanism
for pulsars works in these conditions} Unfortunately, at a rate of
$10^{-5}$ per year in the Galaxy, we only expect 0--1 such events in
our current sample of supernovae.

\section{Core-Collapse Supernovae}
Stars more massive than $\sim$8\,M\sun\/ also end in core-collapse.
During their lives successive stages of nuclear burning build up a
massive iron core in the stellar center.  This iron core is supported
by electron degeneracy and thermal pressures.  When the density and
temperature in the core become so high that iron is dissociated into
alpha particles and electron capture occurs, the support pressure is
suddenly removed and the core collapses.  As it collapses, the core
density and temperature increases, causing more iron dissociation and
electron capture which leads to a runaway infall of the core.  Just as
with AICs, the core collapses until it reaches nuclear densities,
where nuclear forces and neutron degeneracy pressure abruptly halt the
collapse.

Astronomers have long understood that the potential energy released as
a star collapses down to a neutron star could power a supernova
explosion (Baade \& Zwicky 1934).  However, it was not until 1966 that
Colgate \& White realized that neutrinos could be the medium which
transported energy released during the collapse of the core into the
outer layers of the star, which would subsequently explode and drive
the supernova explosion.  Since this time astronomers have continued
to refine the neutrino-driven model.  Indeed, core-collapse supernovae
are one of the few objects in astronomy that astronomers do not invoke
fudge factors to explain (albeit, this means that we do not yet match
the observations all that well).

The basic mechanism behind core-collapse supernovae has developed from
three decades of study and is very similar to AICs.  The main
difference arises from the fact that the proto-neutron star must
somehow eject $\gtrsim 10-15$\,M\sun\/ of material, instead of
0.1\,M\sun\/ as in the case of AICs.  After bounce, the inner portion
of the star rains down upon the proto-neutron star, preventing a
quick, AIC-like explosion.  A convective layer above the proto-neutron
star and below the pressure cap of the infalling material converts the
heat deposited by neutrinos into kinetic energy, aiding the explosion.
Eventually, the energy in the convective layer is sufficient to overcome 
the ram pressure and a supernova explosion is launched.

As the mass of the collapsing star increases, the basic picture
described above of core-collapse supernovae begins to change.  Above
20--25\,M\sun, the supernova explosion is too weak to eject the entire
star, and much of the star ($> 2\,$M\sun) falls back onto the neutron
star 100--100,000\,s after the supernova explosion (and after the GW
emission).  This fallback matter pushes the remnant mass above the
maximum neutron star mass, and it collapses to form a black hole.
Beyond $\sim$40--50\,M\sun\/, the convective layer is unable to
overcome the pressure of the infalling material.  No supernova
explosion is launched and the star collapses to form a black hole.
These direct-collapse stars will differ from normal core-collapse
simulations in both GW emission and optical output.  They are known as
``collapsars'', and constitute one of the favored models for gamma-ray
bursts (Woosley 1993).  Unfortunately, beyond $\sim$40--50\,M\sun,
mass-loss from stellar winds can dramatically change the mass of the
star before collapse, and it may be that nature does not produce any
high-metallicity collapsar progenitors.

\subsection{Formation Rate and Angular Momentum}

The formation rate of core-collapse supernovae is fairly well known,
and lies somewhere between 1 per 50--140 years in the Galaxy
(Cappellaro et al 1997).  5--40\% of these supernovae produce black
holes through fallback accretion (Fryer \& Kalogera 2001).  Because
mass-loss from winds are uncertain, the fraction of massive stars
which collapse directly into black holes is much less well determined
(Fryer \& Kalogera 2001).  And for all core-collapse models, we do not
know the fraction of these massive stars (if any) that are rotating
rapidly enough to emit detectable amounts of GWs.

One way to get a handle on the angular momentum is to study pulsars,
the compact remnants of core-collapse supernovae. From measurements of
young pulsars we know that at least some neutron stars are born with
periods faster than 20\,ms.  But whether or not any neutron stars are
born with millisecond periods is hard to ascertain.  The problem is
that pulsars spin down as they emit radiation, but we don't know
exactly how fast the spin-down occurs.  The most recent analysis by
Chernoff \& Cordes (private communication) found that they could fit
the initial spin periods with a Gaussian distribution peaking at 7\,ms
with sub-ms pulsars lying beyond the 2-sigma tail.  Does this mean
that less than $10\%$ of pulsars are born spinning with millisecond
periods, or does it mean that many pulsars are born spinning rapidly
and GW emission removes a considerable amount of their angular
momentum?  In addition, the analysis of Chernoff \& Cordes is very
sensitive to their choice of spin down rates and other uncertainties
in their population study, and they stress that such results should be
taken with a great deal of caution.

Stellar theorists have now produced models of core-collapse
progenitors which include angular momentum (Heger 1998).  Although
these simulations include a number of assumptions about the angular
momentum transport in the massive star, they give us some handle on
the angular momentum distribution in the collapsing core.  We base our
analysis on the angular momentum profiles from the core-collapse
simulations of Fryer \& Heger (2000), which uses these latest
progenitors and modeled the core-collapse through supernova explosion
(Models 1,5; Fig. 2).  The angular momentum in these collapsing cores
is much less than what is typically used in GW calculations, and we'll
discuss the differences in the results below.

\subsection{Gravitational Waves}

\subsubsection{GW from Collapse or Explosion}

Because the angular momentum distributions used by Fryer \& Heger
(2000) have peak values significantly lower than those used in the
past, there is no centrifugal hang-up.  The collapse proceeds nearly
identically to a non-rotating star, with a density distribution peaked
at the center of the star (Fig. 3).  This makes it harder for bar-mode
instabilities to develop and produces weaker GW emission.  During
bounce, the neutron star is not compact enough to quickly drive
bar-mode instabilities.  However, Fryer \& Heger (2000) found that the
explosion produced by these rotating core-collapse supernovae is much
stronger along the poles than along the equator, causing more of the
low angular-momentum material to be ejected.  Hence, after the
explosion ($\sim 1$\,s after collapse), $\beta$ can increase to high
enough values that bar-mode instabilities are likely to develop
(Fig. 1).  The proto-neutron star extends in all cases beyond
$\sim$1\,M\sun, corresponding to values of $\beta$ which are certainly
above the secular instability limit and probably above the dynamic
instability limit (see \S 2.2).  Notice in Fig. 1 that $\beta$ is
actually larger for the model which has less angular momentum.  This
is because this model has contracted more and is spinning more
rapidly.

Numerical evaluation of the quadrupole formula (\S 2.1) predicts a
peak strain of $4.1 \times 10^{-23}$ at 10\,Mpc and at $f \sim 20$ Hz.
The total energy released in GWs is about $2\times10^{44}$\,ergs.
These results are comparable to those for AICs.  However, since
core-collapse supernovae are nearly 1,000 times more common than AICs,
we are much more likely to detect a nearby core-collapse supernovae.
To get a better handle on the potential of bar instabilities in
core-collapse, we use our collapse models to construct a bar and
calculate the GW emission from this bar (\S 2.3).  We do this
calculation by assuming that all of the matter up to some enclosed
critical mass becomes unstable and forms a bar (conserving angular
momentum), and then calculate the GW emission as a function of total
unstable mass (Fig. 6).  Note that strains as high as $10^{-22}$ are
possible with frequencies as high as a kHz (Table 1).  However, one
should keep in mind we have assumed that {\it all} of the enclosed
mass ends up in the bar.  These estimates are relatively strong upper
limits (although the strain could increase if we allowed the bar to
contract and spin up).  We illustrate the detectability of waves from
bar instabilities in Figure {\ref{fig:barmode_15}}.  Each point on
this plot illustrates a different possible bar, varying the amount of
mass that participates in the instability.  Open circles illustrate
the wave strain for a single GW cycle; filled circles give the
characteristic strain obtainable if the bar emits coherently for 100
cycles (see Appendix A for further discussion).  This plot
demonstrates that bar modes are potentially promising sources of waves
if bars remain coherent for at least a moderate ($\sim 50$--$100$) number of GW
cycles.

As the density is centrally peaked, the fragmentation instability is
unlikely to occur in core-collapse supernovae.  However, if it did
occur, the strain would be $\sim 10^{-22}$, at frequencies as high as
2 kHz (Table 1).

\subsubsection{Remnants}

Core-collapse supernovae produce both neutron stars and black holes.
GW emission from young neutron stars produced in core-collapse
proceeds similarly to AICs.  The major difference is that in
core-collapse supernovae a considerable amount of fallback can occur
at late times, subsequently spinning up and reheating the young
neutron star.  Generally, lower mass stars produce less fallback, but
the fallback material accretes at a higher rate and at earlier times
(Fryer et al. 1999b).  We have calculated the GW emission, spin down
rate, and r-mode amplitude for two fallback rates from Fryer et
al. (1999b): 0.003\,M\sun\,s$^{-1}$ between 20--50\,s (dotted line;
Fig. 4) after the launch of the explosion, and 0.0003\,M\sun\,s$^{-1}$
between 2,000--9,000\,s after the explosion (thick dashed line; Fig. 4).
Note that fallback can cause the neutron star to spin up fast enough
to emit a second burst of GWs.  In all cases, the total energy emitted
in GWs exceeds $10^{52}$\,ergs, with a maximum power of over
$10^{50}$\,ergs\,s$^{-1}$. Fallback is not the only
way to produce multiple bursts of gravitational waves.  Numerical
simulations have shown that the r-mode amplitude can grow above one and
then dissipate, followed by additional growth (Lindblom, Tohline,
\& Vallisneri 2001).  It may be difficult to tell whether a second
burst is due to fallback or simply regrowth after dissipation.

As with the AICs, we stress that if the neutron star magnetic field is
less than $\sim 10^{13}$\,G, there will be no observational evidence
of the high spin periods until late times.  After a year, the GW
emission will have spun the pulsar down to 11\,ms, which roughly
matches pulsar observations (Chernoff \& Cordes, private
communication).  If the magnetic field is higher, these fast-spinning
pulsars will affect the dynamics of the exploding material.  Could
this explain the polarization measurements of supernovae?  If this is
truly the mechanism which causes supernovae to be polarized, some
supernovae should be unpolarized (e.g., the progenitor of the
relatively low-field Crab pulsar).

For stars more massive than 25\,M\sun\/ it is likely that a large
amount of material will fall back onto the newly-formed neutron star,
causing it to collapse to a black hole.  The ringing of this newly
formed black hole will produce GW emission (\S 2.5).  In the most
rapid fallback cases, the accretion rate onto the black hole can be as
high as 0.01\,M\sun\,s$^{-1}$ (Fryer et al. 1999b; MacFadyen, Woosley,
\& Heger 2001).  For our calculations, we assumed that the 
black hole formed with a mass of 3\,M\sun and then accreted at a rate
of 0.01\,M\sun\,s$^{-1}$ for 500\,s, holding the spin fixed at
$a/M_{\rm BH}=0.7$ for the duration of the accretion process.  For an
extremely optimistic estimate of the GW signal from black hole
ringing, we assume that the material accretes in 0.5\,s clumps
($T_{\rm thump} =0.5\,$s), and has a very high radiation efficiency
($\varepsilon = 0.5$).  Even with such optimistic parameters, the
strain for a 10\,Mpc supernova turns out to be less than
$5\times10^{-24}$ and is at frequencies above 2\,kHz, both too weak
and outside the most sensitive band of LIGO-II.  The mode mixing
parameter $\alpha_{\rm ring}$ has little impact on this result.

For stars more massive than 50--60\,M\sun, it is likely that the star
does not form a supernova at all, and instead collapses directly to a
black hole.  These collapsars are likely to accrete at rates as high
as 1--10\,M\sun\,s$^{-1}$.  For our collapsar model, we again assume
the black hole forms with an initial mass of 3\,M\sun\/ and then
accretes at a rate of 1\,M\sun\,s$^{-1}$, which decreases to
0.1\,M\sun\,s$^{-1}$ after 3\,s and stays constant for roughly 20\,s.
The black hole is assumed to start with an $a/M_{\rm BH}=0.5$ and
spins up as it accretes to 0.7 in 1\,s and then up to 0.86 in the
remaining time.  This model was picked to mimic the results of
MacFadyen \& Woosley (1999).  We calculated two models, assuming the
ringing is dominated by the $m=0$ and $m=2$ modes respectively.  At $z
= 1$ and using $\varepsilon = 0.5$ and $T_{\rm thump} = 0.5\,$s, the
strain can be as high as $2 \times 10^{-23}$ at frequencies around
2\,kHz (Fig. 8).  Unfortunately, the sensitivity of LIGO type
detectors is generally poor at such frequencies.  Coupled with the
fact that our choices for $\varepsilon$ and $T_{\rm thump}$ are
extremely optimistic, we believe it is extremely unlikely that the
ringing waves from black holes formed in collapsars will be seen by GW
detectors (Fig. 8).

\section{Collapse of Very Massive Stars}

At solar metallicity, stellar winds severely limit the pre-collapse
mass of massive stars, and very few massive stars will remain massive
up to the time of collapse.  These winds are driven by the opacity of
metals in the stellar envelope.  It is likely that as we reduce the
fraction of metals in the star, mass-loss from winds will decrease.
Population III stars are the first generation of stars formed in the
early universe, when virtually no metals existed (stars produce all of
the metals we see today).  In this section we review the death of very
massive, population III stars (100--500\,M\sun).  Like
Chandrasekhar-massed white dwarfs, these stars must suffer one of two
fates: either they explode in a giant thermonuclear explosion
(``hypernova'') or they collapse to form black holes.  The fate is
determined by the stellar mass.  If the star's mass exceeds
$\sim$260\,M\sun, it will collapse to a black hole (Fryer et al. 2001;
Baraffe, Heger, \& Woosley 2001). However, if the star is rotating,
rotational (plus thermal) support prevents the star from undergoing
immediate collapse to a black hole (Fryer et al. 2001).  Rotating, 
very massive stars collapse and bounce, forming a much larger compact 
core than those produced by core-collapse supernovae:  a 50--70\,M\sun, 
1000--2000\,km proto-black hole instead of the 1\,M\sun, 100\,km 
proto-neutron star.  This rotating proto-black hole is susceptible to bar
instabilities and may produce a strong GW signal (see also Madau \&
Rees (2001)).

\subsection{Formation Rate and Angular Momentum}

Estimating an accurate rate of core-collapse from very massive stars
depends on two rather uncertain quantities: the amount of matter found
in population III stars and the number of these stars which actually
collapse to form black holes.  The mass distribution of stars at birth
is known as the initial mass function (IMF).  Today the IMF is peaked
toward low mass stars, such that 90\% of stellar core-collapse occurs
in stars between 8 and $\sim$20\,M\sun, and only 1\% of core-collapse
occurs in stars more massive than 40\,M\sun.  However, it has long
been believed that the first generation of stars after the Big Bang
tended to be more massive than stars formed today (e.g., Silk 1983;
Carr \& Rees 1984).  Recent simulations by Abel, Bryan, \& Norman
(2000) suggest that the typical mass of first
generation stars is $\sim 100$\,M\sun\/ and it could be that a
majority of Population III stars had masses in excess of 100\,M\sun\/.

The light from these very massive stars re-ionizes the early universe,
and from this we can derive a constraint on the formation rate of
these stars.  Although we expect that these photons ionized a
significant fraction of the early universe, there should not be so
many stars that they ionize the universe several times over.  Using
our best estimates of the re-ionization fraction, the amount of
ultraviolet photons produced by these massive stars, and the
ionization efficiency of massive stars, one estimates that roughly
$0.01\%-1\%$ of the baryonic matter in the universe was incorporated
into very massive stars (Abel et al. 2000).  This corresponds to
roughl $10^4 - 10^7$ very massive stars produced in a
$10^{11}$\,M\sun\/ galaxy, or a rate of massive stellar collapse as
high as one every few thousand years

We should temper these results with two conditions.  First, these
stars are Population III stars, and so are born at high redshift
($z\gtrsim 5$). As they evolve to collapse in less than a few million
years (Baraffe et al.  2001), they will only be observed at the high
redshifts of their birth.  Although we might believe our formation
rate of very massive stars (within a few orders of magnitude), it is
currently impossible to determine how many very massive stars are
produced with masses beyond the $\sim 260$\,M\sun\/ mass limit
necessary for black hole formation.  The Galaxy could produce a
million of these objects, or perhaps just a few hundred. Assuming
1--10 million very massive stars per galaxy beyond $z = 5$ gives us a
secure upper limit.  The rotation of these stars has again been
calculated using the stellar evolution code developed by Heger (1998),
and for this analysis we use the Fryer et al. (2001) rotation profiles
(Fig. 2).

\subsection{Gravitational Waves}

\subsubsection{GW from Collapse or Explosion}

The proto-black hole formed in the collapse of a massive
star is expected to become secularly unstable (Fig. 1), and
these secular instabilities are likely to develop before the
proto-black hole collapses to a black hole (Fryer et
al. 2001).  With the large amount of mass ($\sim 70$\,M\sun)
and angular momentum (Fig. 2), it is not surprising that
these objects produce strong GW signals.  However, the
cosmological redshift moves the peak of the source waves out
of the band of LIGO detectors: even at the relatively low
value $z = 5$ (luminosity distance $\sim48$\,Gpc), the
strain from bar modes peaks at frequencies less than 10 Hz,
with a strain $8 \times 10^{-23}$.  This is well below the
LIGO II threshold.  Even coherent integration over $\sim100$
cycles is unlikely to produce a detectable signal; see
Figs. 6 and 9.

The waves from massive star collapse may be detectable if a
fragmentation instability occurs.  With our crude model of
fragmentation, we find that both strain and frequency are boosted if
the core splits into two pieces which then fall into a Keplerian
orbit, conserving angular momentum.  If this instability occurs and
the pieces orbit coherently for $\sim 10$ cycles, these waves may be
detectable at redshifts $z\sim5$; see Figs. 9 and 10.

\subsubsection{GW from Remnants}

Because the proto-black hole collapses before it can cool, the bulk
viscosity prevents the growth of r-modes up to collapse.  No GW
radiation will occur from r-modes during the collapse of very massive
stars.

After the black hole forms out of the inner 10--20\,M\sun, material
rapidly accretes onto it (Fig. 11).  Like collapsars, this accretion
rate can be very high and produce strong ringing in the newly formed
black hole.  Using the accretion rate and spin evolution from the
simulations of Fryer et al.\ (2001), we estimated GW emission from $m
= 0$ and $m = 2$ modes.  We set the radiative efficiency parameter
$\varepsilon = 0.1$ (moderately optimistic) and put $T_{\rm thump} =
0.1\,$s (indicating a relatively clumpy flow).  At $z = 5$, the waves
are in LIGO's band, but do not achieve strains large enough to
guarantee detection: we find that $h \sim 6\times 10^{-23}$ initially
(nearly touching the LIGO-II noise level), and that the strain rapidly
decreases after that; see Fig.\ {\ref{fig:ring}}.  We also illustrate
these waves for a collapse at $z = 20$; they are at even lower
frequency and weaker strain.  This figure makes us rather pessimistic
about the likelihood of measuring black hole ringing waves from the
collapse of very massive stars.  We note, however, that in the case $z
= 5$ the accumulated signal-to-noise is of order unity.

\section{Conclusion}

We have examined GW production under a wide range of stellar collapse
scenarios, assuming various mechanisms for GW generation. Our results
indicate that some of these waves are likely to be of detectable
strength, even though modern models of collapse progenitors predict
less angular momentum than they did previously.  The main results of
our analysis are as follows.

For the accretion-induced collapse (AIC) of a white dwarf, the largest
potential source of GWs during the explosion is likely to be from bar
formation.  However, the data used in this analysis indicate that the
AIC proto-neutron star will not be unstable to bar formation, so we
have not considered this mechanism.  (We note, though, that Liu \&
Lindblom (2001) consider white dwarfs with larger angular momentum,
and these would certainly be unstable to bar mode formation; further
work will no doubt clarify the effectiveness of such objects for GW
production.)  The newly formed remnant is hot enough that it should be
unstable to the production of r-mode GWs.  The event rate for AICs is
fairly low ($\sim 10^{-5}\,$yr$^{-1}$ per galaxy), so one must
consider events as far out as 100 Mpc.  At this distance, both r-mode 
and bar-mode (if they occur) waves are unlikely to be detectable.

For core-collapse supernovae, a bar-mode instability is likely to
develop shortly after the launch of the supernova shock, similar to
the case of AICs.  Core-collapse supernovae are much more frequent
than AICs ($\sim 10^{-2}\,$yr$^{-1}$ per galaxy), so we do not have to
look quite so far out; we consider collapse out to about 10 Mpc.  For
a range of bar formation scenarios, we find frequencies and strains
that fall above the LIGO-II noise curve.  High signal-to-noise
measurements are possible if the bar remains coherent for tens to
hundreds of cycles.  The collapse forms a rapidly spinning neutron
star, which may be unstable to r-mode emission.  The r-modes can form
in the hot remnant itself, or be ``reactivated'' somewhat later when
hot material falls back onto the star, reheating it and increasing its
spin.  These waves should be detectable by LIGO-II.  Very massive
progenitors form a black hole instead of a neutron star.  This case
produces no r-mode emission, but there will be GWs from ringing of the
black hole horizon.  However, even with ridiculously optimistic
assumptions about the nature of the matter flow onto the hole, we find
that these waves are completely undetectable --- their strains are
very weak, and they are emitted at high frequencies ($f \sim
2000$--$5000$ Hz) where detectors have poor sensitivity.

We also consider the collapse of 300\,M\sun\/ stars as a source of GW
emission.  Although these massive stars may have been quite common in
the first generation of stars, it is unlikely that they formed at
redshifts below five.  When very massive stars collapse, they pass
through a phase where the matter is densely clumped into a proto-black
hole.  This rapidly rotating object is susceptible to bar mode
instabilities, producing a fairly large GW strain. However, the GWs
are unlikely to be detectable because of the large frequency shift due
to the cosmological redshift and we find that bar formation at
redshift $z\sim5$ is marginally detectable only if the bar persists
for hundreds of cycles.  If the proto-black hole material were to
fragment into clumps (which we model as a binary), it could yield a
signal detectable by LIGO-II.  As the proto-black hole remains
extremely hot until collapse, the bulk viscosity prevents an r-mode
instability from developing.  Although we believe it is unlikely, it
is possible that the ringing of the newly born black hole {\it
might} be detectable.  We find that if the star collapses at $z \sim
5$, the ringing waves are emitted at high enough frequency ($f
\sim 30$--$50$ Hz) that detector noise may not overwhelm the signal.
If the collapse is asymmetric and ``clumpy'' enough to severely
distort the hole for several seconds, the ringing waves are likely to
be of interesting strength.  We hope that this result will motivate
more careful analyses of ringing waves from BHs that are produced in
massive stellar collapse (an example of which is described in
Papadopoulos \& Font (2001)).  It is also to be noted that Madau \&
Rees (2001) have recently examined massive black holes as remnants
from population III stars.  In their paper they mention the emission
of GWs from the capture of such holes onto supermassive black holes at
the centers of galaxies.  Such waves are very different from those
discussed here, and in fact would be sources for low-frequency
space-based detectors such as LISA.  We have not considered this
mechanism here.

For all of our estimates we have made a number of assumptions, chosen
to ensure a safe maximum of the gravitational wave signal.  Therefore
all of our results should be taken as upper limits, and it is
certainly possible that the GW signals are weaker than
our calculations suggest.  In addition, although based on recent
collapse calculations, any estimate of the gravitational wave
signal will be limited by the uncertainties in the core-collapse 
models.  However, it is unlikely that our upper limits will
change significantly as collapse simulations improve.

\acknowledgements

C.L.F. gratefully acknowledges the support of the Institute for
Theoretical Physics where this work was initiated.  The authors are
grateful to Luciano Rezzolla and John Beacom for useful comments, to
David Shoemaker and Kip Thorne for pointing out some errors, and
to Warner Miller for interesting discussions in the course of this
work.  The work of C.L.F was funded by the ITP visitor program and a
Feynman Fellowship (LANL).  D.E.H and S.A.H. are supported by NSF
Grant PHY-9907949 at the ITP.

\appendix

\section{Detectability criteria}
\label{app:detect}

\setcounter{equation}{0}

Because of the weakness of GWs, sophisticated data processing methods
are needed to pull astrophysical signals from a detector's noisy data
stream.  Here we discuss some useful figures of merit that
characterize the effectiveness of such techniques.  We use these
results to describe how well the GW strains discussed in this paper
can be detected by GW detectors, focusing in particular on the second
generation LIGO interferometers (``LIGO-II'').

One statistic in particular is of great importance to us: the signal
power $\rho^2$, given by
\begin{equation}
\rho^2 = 4\int_0^\infty df {|\tilde h(f)|^2\over S_h(f)}\;,
\label{eq:mfsnr}
\end{equation}
where ${\tilde h}(f)$ is the Fourier transform of the
gravitational waveform $h(t)$, and $S_h(f)$ is the spectral
density of strain noise.

One of the most important methods of searching for GWs is called
``matched filtering''.  This method uses a model or template for the
waveform to construct a linear filter.  The instrumental data are then
correlated with the filter, yielding a signal-to-noise ratio (SNR).
If a signal in the data stream matches the template, then the SNR has
a mean value $\rho$:
\begin{equation}
\left({S\over N}\right)_{\rm MF} = \rho\;.
\end{equation}
The signal power is often referred to as the matched
filtering SNR.  It is simple to prove that this SNR is the
maximum obtainable with linear filters, so this technique is
sometimes called ``optimal filtering''.  Signals are
detectable if the matched filtering SNR exceeds a threshold,
$\rho_{\rm thresh}$, whose value depends on a large number
of parameters (number of templates in the filter bank,
duration of the templates, number of detectors, etc.).  As a
rough rule of thumb, $\rho_{\rm thresh} \simeq 5$.  Further
discussion of this point can be found in Flanagan \& Hughes
(1998), particularly the text near Eqs.~(2.8) and~(2.9).

Because the GWs from a particular source depend upon the source's position on
the sky, and its orientation, the SNR should be averaged over these
quantities.  Flanagan \& Hughes (1998) have shown that the matched filtering
SNR, so averaged, depends only on the spectrum of GW energy $dE/df$:
\begin{equation}
\langle\rho^2\rangle = {2(1+z)^2\over5\pi^2D(z)^2}\int_0^\infty
df\,{1\over f^2 S_h(f)}{dE\over df}\left[(1 + z)f\right]\;,
\label{eq:snrsqr_ave}
\end{equation}
where $z$ is the cosmological redshift of the source, and
$D(z)$ is its luminosity distance.

The matched filtering SNR formula can be rewritten as follows:
\begin{equation}
\langle\rho^2\rangle = \int_0^\infty
{df\over f}\left[{h_{\rm char}(f)\over h_{\rm noise}(f)}\right]^2\;.
\end{equation}
Using Eq.\ (\ref{eq:snrsqr_ave}) as our model, the ``characteristic
signal strain'' $h_{\rm char}(f)$ is given by
\begin{equation}
h_{\rm char}(f) = {\sqrt{2}(1+z)\over\pi D(z)}\sqrt{{dE\over
df}[(1+z)f]}\;.
\label{eq:hchar}
\end{equation}
Note that if one knows the characteristic strain for a single cycle,
the value for $N$ cycles is just
\begin{equation}
h_{\rm char} \simeq \sqrt{N} h_{\rm char,\ 1\ cycle}
\end{equation}
since $dE/df$ accumulates with the number of cycles.  (The equality is
exact if the signal is monochromatic.)  The ``mean noise strain''
$h_{\rm noise}(f)$ is
\begin{equation}
h_{\rm noise}(f) = \sqrt{5fS_h(f)}\;.
\label{eq:hnoise}
\end{equation}
The factor of $\sqrt{5}$ in this equation arises from the detector's
sensitivity pattern, which effectively increases the noise for sky
averaged sources.  Note that the noise spectral density $S_h(f)$ used
in Eq.\ (\ref{eq:hnoise}) is the square of the quantity plotted in
Gustafson et al (1999).  The strains $h_{\rm char}(f)$ and $h_{\rm
noise}(f)$ indicate in a simple way the relative strength of the
astrophysical GW signal and the noise at a particular frequency.
Comparing them is an effective way of quickly estimating the signal
power: if $h_{\rm char}(f) > h_{\rm noise}(f)$, then the signal is
likely to be detectable with matched filtering, and hence worth
further analysis.  In Figs.\ 5 and 7--9 we show $h_{\rm char}(f)$ for
the various GW emission scenarios discussed in this paper.

Throughout this paper we use the broad-band LIGO-II noise curve
described in {\cite{whitepaper}}.  In particular, it is used to
compute $h_{\rm noise}(f)$ as plotted in the figures.  LIGO-II is a
planned upgrade to the detectors at the two LIGO sites.  Research and
development for various LIGO-II components is in progress, with final
implementation set to begin in 2006 or shortly thereafter.  One of
LIGO-II's design goals is to have a moderately tunable sensitivity
profile; the broad-band curve we use here is a useful configuration
for analyzing the detectability of sources that are not very well
understood.  Other possible configurations include a curve optimized
for detecting binary neutron stars, and a curve with moderately narrow
band sensitivity.  See {\cite{whitepaper}} for further discussion and
details.

A second statistic that could easily be computed (though we do not do
so here) is the time-frequency volume ${\cal N}$:
\begin{equation}
{\cal N} = 2 T \Delta f \;.
\label{eq:voldef}
\end{equation}
The quantity $T$ is a duration and $\Delta f$ a bandwidth.  For the
purpose of estimating the detectability of astrophysical signals, $T$
is the duration of the signal and $\Delta f$ its frequency bandwidth.
(In an actual GW search, $T$ and $\Delta f$ may be treated as
variational parameters so that events are not missed.)

This statistic is useful because it helps to understand how well
techniques that are not based on matched filtering might perform.  A
large value of ${\cal N}$ indicates that the signal power is
``smeared'' over a wide range of frequencies or Fourier bins, and that
a signal may be difficult to reconstruct without a great deal of prior
information.  For example, suppose we know that the signal has support
only in some bandwidth $\Delta f = f_{\rm high} - f_{\rm low}$.
Rather than using a matched filter (which we do not have enough
information to construct), data analysis would use a simple band-pass
filter.  This filter chops out all power in the datastream at
frequencies above $f_{\rm high}$ and below $f_{\rm low}$.  The SNR
obtained with this filter is the ratio of the signal peak amplitude to
the noise peak amplitude in the filtered datastream.  Following the
analysis of Flanagan \& Hughes (1998), this is given by
\begin{equation}
\left({S\over N}\right)_{\rm BPF} \simeq {\rho\over\sqrt{\cal N}}\;.
\label{eq:BPF_to_MF}
\end{equation}
A useful rule of thumb is that signals are detectable with band-pass
filtering if $(S/N)_{\rm BPF}\ge 1$.

The band-pass filtering SNR is a kind of worst case scenario,
applicable when very little is known about the signal's
characteristics.  The matched filtering SNR is a best case scenario,
for when the signal is very well understood.  A more
middle-of-the-road diagnostic is the excess power statistic, first
introduced in Flanagan \& Hughes (1998), and discussed at length in
Anderson et al. (2001).  This search technique requires knowledge of the
duration $T$ and frequency bandwidth $\Delta f$.  It works by dividing
the frequency domain data into bins of width $\delta f \sim 1/T$ and
incoherently combining the power in each bin.  The resultant
statistic, ${\cal E}$, is a measure of the total energy in the
datastream.  Detection of a signal occurs when the value ${\cal E}$
exceeds what would be expected from noise alone in a statistically
significant way.  The analysis of Anderson et al. (2001) tells us that
this occurs when
\begin{equation}
\rho^2 > \sqrt{2{\cal N}}\;.
\label{eq:excesspower_detcrit2}
\end{equation}

We thus have some useful rules of thumb for estimating the
detectability of a GW signal.  First, calculate the sky-averaged SNR,
Eq.\ (\ref{eq:snrsqr_ave}), and the time frequency volume, ${\cal N} =
2 T\Delta f$.  If $\langle\rho^2\rangle^{1/2}\equiv\rho_{\rm ave}$ is
of order 5 or so, then this is a potentially interesting source.  If
$\rho_{\rm ave}$ is closer to unity, then it will be hard to detect,
but probably should not be dismissed out of hand.  On the other hand,
if $\rho_{\rm ave}$ is much less than unity there is little chance the
signal will be detected.

If $\rho_{\rm ave}$ indicates that the source is interestingly strong,
examine $\rho_{\rm ave}/\sqrt{\cal N}$, the band-pass filtering SNR.
If this is of order 1 or greater, then the source should be detectable
without too much sophisticated data analysis.  If $\rho_{\rm ave}$ is
large but $\rho_{\rm ave}/\sqrt{\cal N}$ is small, then some detailed
knowledge of the source's characteristics will be needed to aid data
analysis.  A good idea of how much detail is needed can be obtained by
checking the number $\rho_{\rm ave}/{\cal N}^{1/4}$.  If this measure
is large, then the excess power statistic is likely to be useful.  In
this case, just knowing the signal's likely bandwidth and duration
should be enough to significantly aid data analysis.

\begin{deluxetable}{llccccc}
\tablewidth{38pc}
\tablecaption{Gravitational Wave Emission}
\tablehead{ \colhead{Object} & 
\colhead{Emission} & 
\colhead{Rate\tablenotemark{a}} &
\colhead{Typical} 
& \colhead{h(f)}
& \colhead{f(Hz)} & \colhead{Max. Power} \\
\colhead{} & \colhead{Mechanism} & \colhead{(yr$^{-1}$)} 
& \colhead{Distance} & \colhead{($10^{-24}$)} & \colhead{} &
\colhead{ergs\,s$^{-1}$}}

\startdata

AIC\tablenotemark{b} & & $\lesssim 10^{-5}$ & 100\,Mpc & & & \\
 & Numerical\tablenotemark{c} & & & $5.9$ & $\sim 50$ 
& $10^{48}$ \\
 & r-modes\tablenotemark{c} & & & $0.35$ & $\sim 1000$ 
& $10^{50}$ \\

SN\tablenotemark{b} & & $\lesssim 10^{-2}$ & 10\,Mpc & & & \\
 & Numerical\tablenotemark{c} & & & $41$ & $\sim 20$ 
& $10^{45}$ \\
 & Bar-modes\tablenotemark{c} & & & $100$ & $\sim 1000$ 
& $10^{53}$ \\
 & Binary\tablenotemark{c} & & & $100$ & $\sim 2000$ 
& $10^{54}$ \\
 & r-modes\tablenotemark{c} & & & $0.35$ & $\sim 1000$ 
& $10^{50}$ \\
 & BH Ringing\tablenotemark{c} & & & $5,2$ 
& $\sim 2000,\sim 2000$ & $10^{55}$ \\

300M\sun\tablenotemark{b} & & $\lesssim 10^{7}$ & z=5,z=20 & & & \\
 & Numerical\tablenotemark{c} & & & $80,20$ & $\lesssim 2-10, \lesssim 0.5-2.5$ 
& $10^{42}$ \\
 & Bar-modes\tablenotemark{c} & & & $7,5$ & $\sim 10$ 
& $10^{54}$ \\
 & Binary\tablenotemark{c} & & & $120,80$ & $\sim 30$ 
& $10^{56}$ \\
 & BH Ringing\tablenotemark{c} & & & $60,30$ & $<70,<20$ & $10^{57}$ \\

\enddata

\tablenotetext{a}{For AICs and Supernovae, we have given the rate in 
number per year for a Milky-Way massed galaxy.  The number given 
for the collapse of 300M\sun stars is the rate in the universe per 
year (assuming upper limits on the rate and roughly $10^{10}$ Milky-Way 
massed galaxies in the universe).  Although this number is extremely 
high, remember that only those that occur at low redshifts will even 
rise above the sensitivity curve of LIGO-II.  In addition, the rate is 
only an upper limit of an extremely uncertain number which may be 
many orders of magnitude lower.}
\tablenotetext{b}{AIC $\equiv$ accretion-induced collapse
(see \S 3); SN $\equiv$ core-collapse supernovae 
(see \S 4), note that there are two values for BH ringing, the former 
for fallback, and the latter for collapsars; 300M\sun
$\equiv$ 300\,M\sun\/ population III stars\/ (see \S 5),
note that we consider two values for 300\,M\sun\/ stars, the
former for systems at $z=5$, the latter for $z=20$.}
\tablenotetext{c}{Numerical $\equiv$ Numerical Quadrupole Method (see \S 2.1), 
Bar-modes (see \S 2.2), Binary (see \S 2.3), r-modes (see \S 2.4), BH 
ringing (see \S 2.5).}

\end{deluxetable}

\begin{figure}
\plotfiddle{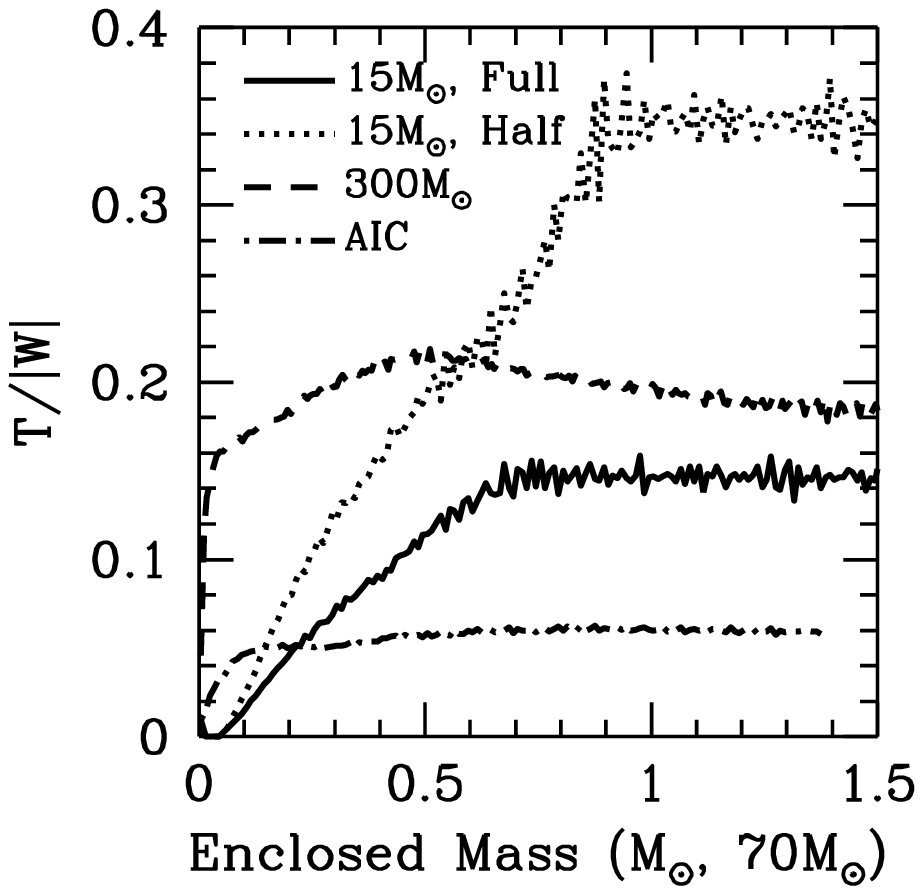}{7in}{0}{90}{90}{-180}{-10}
\label{fig:beta}
\caption{Rotational energy divided by gravitational energy ($T/|W|$)
versus mass for collapsing stars. AIC, 0.18\,s after collapse; rotating
core-collapse (full rotation), 1.6\,s after bounce; core-collapse (half
rotation), 1.4\,s after bounce; 300\,M\sun\/ direct collapse, 1.9\,s after
collapse.  For the core-collapse stars, $T/|W|$ is actually higher for
the star initially spinning at half the rotation rate, as
the system is more compact.}
\end{figure}
\clearpage

\begin{figure}
\plotfiddle{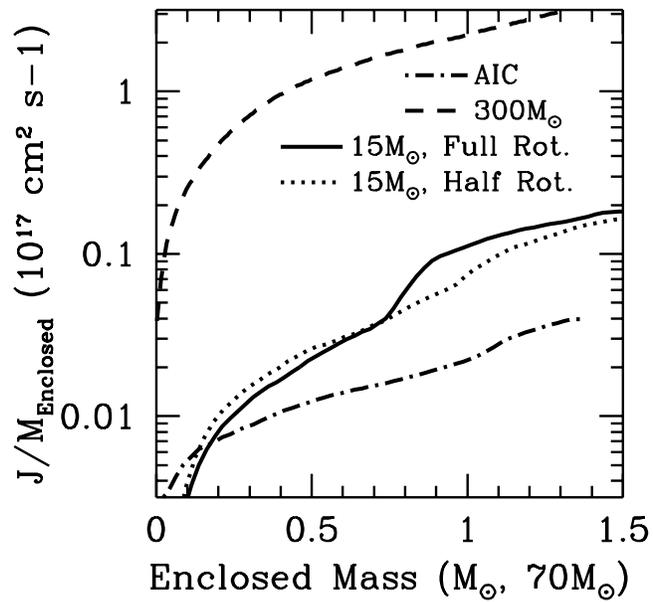}{7in}{0}{90}{90}{-180}{-10}
\caption{Specific angular momentum versus mass for 
collapsing stars. AIC, 0.18\,s after collapse; rotating core-collapse
(full rotation), 1.6\,s after bounce; core-collapse (half
rotation), 1.4\,s after bounce; 300\,M\sun\/ direct
collapse, 1.9\,s after 
collapse.  The 300\,M\sun\/ has, by far, the highest angular
momentum.}
\end{figure}
\clearpage

\begin{figure}
\plotfiddle{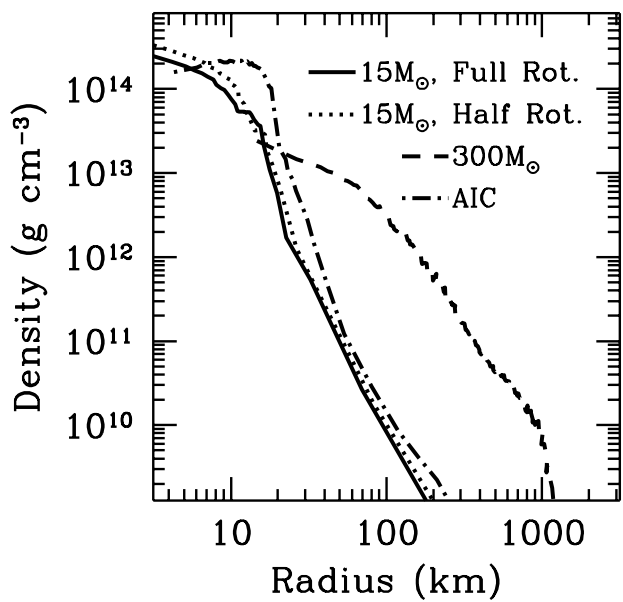}{7in}{0}{90}{90}{-180}{-10}
\label{fig:rhovsr}
\caption{Density versus radius for collapsing stars. AIC, 0.18\,s after 
collapse; rotating core-collapse (full rotation), 1.6\,s after bounce; 
core-collapse (half rotation), 1.4\,s after bounce; 300\,M\sun\/ direct 
collapse, 1.9\,s after collapse.  For the core-collapse simulations, 
the slower rotator is more dense.  Although the maximum density of the 
300\,M\sun\/ direct collapse is much lower than the other core-collapses, 
its mass out to 1000\,km is 50 times that of the other collapsed objects.}
\end{figure}
\clearpage

\begin{figure}
\plotfiddle{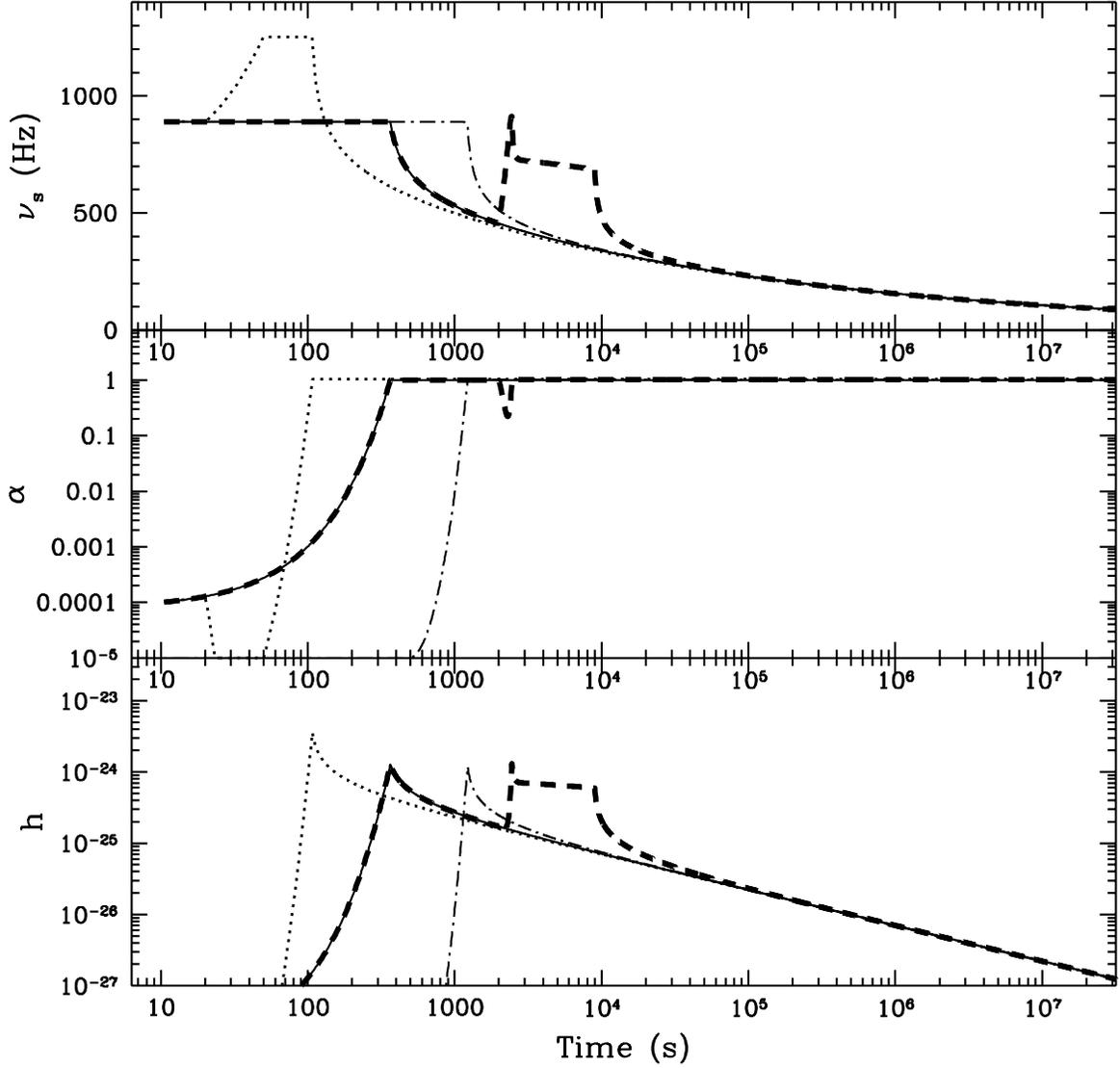}{6in}{0}{80}{80}{-240}{-100}
\label{fig:rmcalc}
\caption{Evolution of the neutron star spin frequency ($\nu_{\rm s}$),
r-mode amplitude ($\alpha$), and strain ($h$) at 10\,Mpc as a function
of time.  The thin solid line corresponds to a simulation assuming no
fallback accretion, with the temperature of the neutron star $T_9$
decaying as $(t/1\,\mbox{yr})^{-1/6}$.  The thin dot-dashed line
corresponds to a temperature decaying as $(t/1\,\mbox{yr})^{-1/3}$.
In core-collapse supernovae it is likely that some fallback will
occur, and the dot-dashed line and thick dashed line correspond to an
accretion rate of 0.01\,M\sun\,s$^{-1}$ 20\,s after the explosion and
0.001\,M\sun\,s$^{-1}$ 2000\,s after the explosion, respectively.
Note that in late-time accretion the fallback material heats up
the neutron star, causing the viscosity to rise and initially damp the
r-mode oscillations.}
\end{figure}
\clearpage

\begin{figure}
\plotfiddle{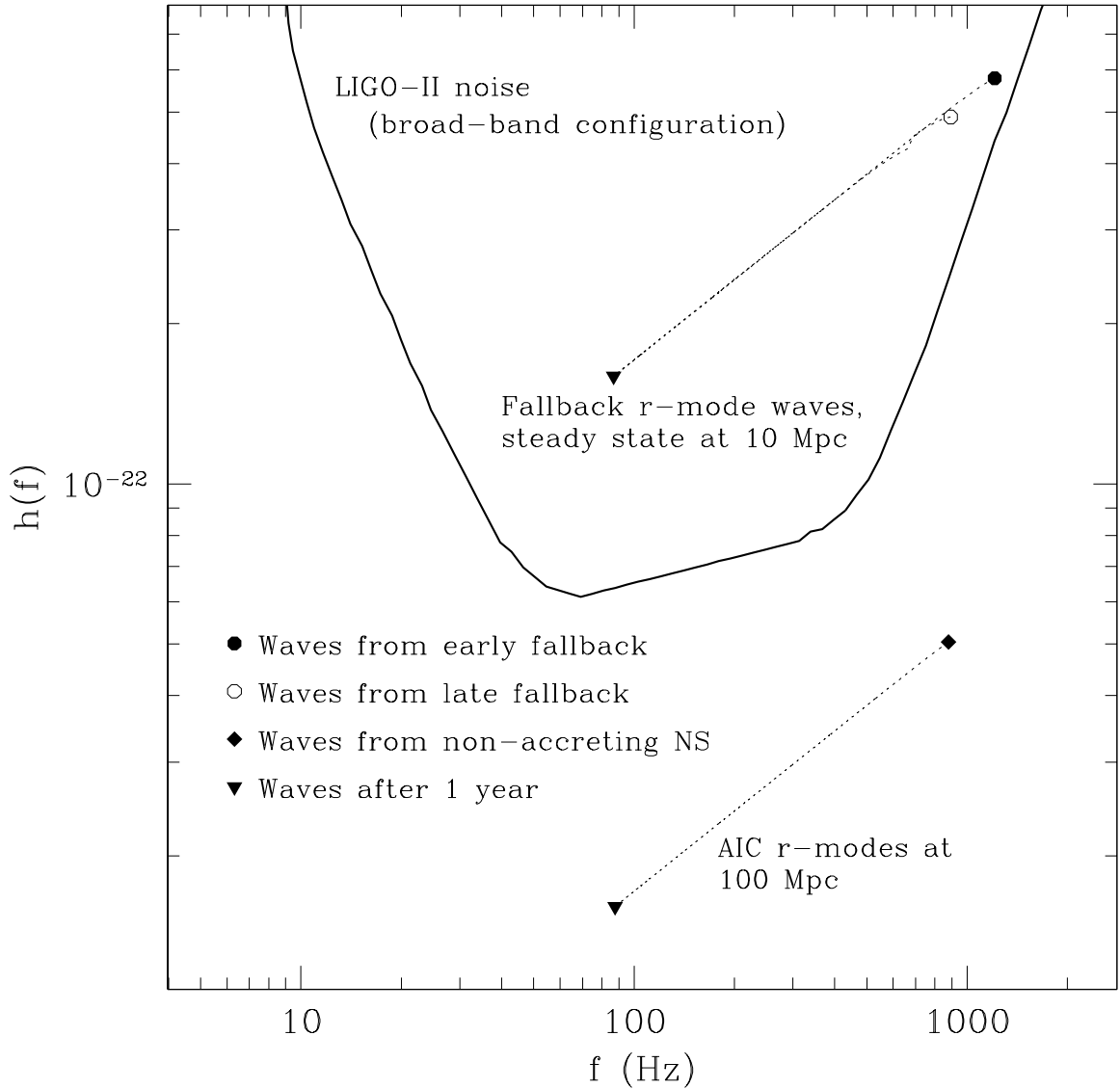}{6in}{0}{80}{80}{-240}{-100}
\caption{
\label{fig:rmode}
Comparison of the characteristic strains associated with r-modes to
the mean noise in enhanced LIGO interferometers.  The top track shows
the waves emitted after early and late fallback reheats and spins up a
newly born neutron star.  The lower track gives the same information
but for a neutron star created in accretion induced collapse.  We show
the AIC waves at 100 Mpc since such events are probably about 1000
times rarer than core-collapse supernovae.}
\end{figure}

\begin{figure}
\plotfiddle{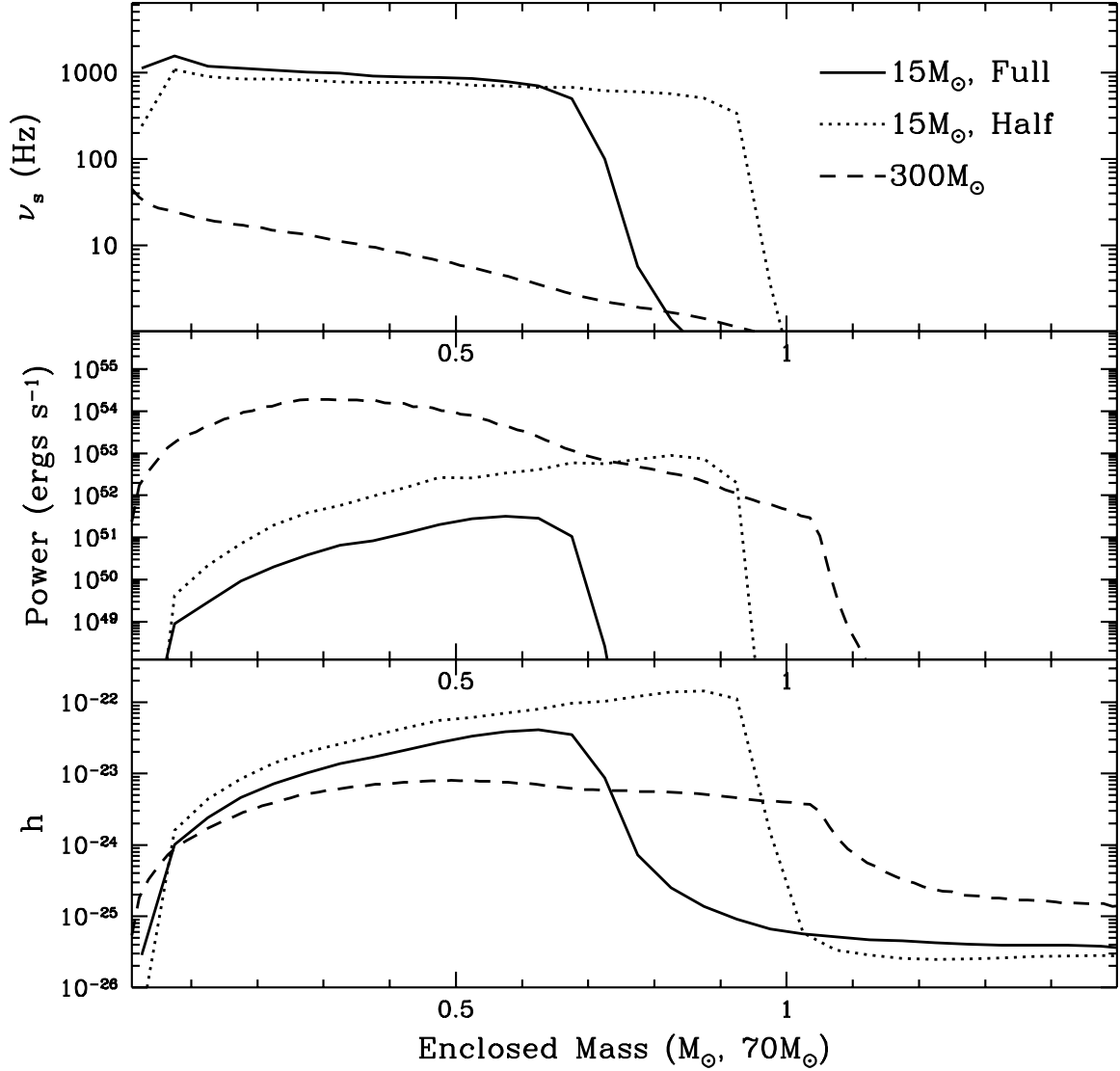}{6in}{0}{80}{80}{-240}{-100}
\caption{Bar-mode spin period, power, and strain versus the mass
encompassed by the bar-mode instability.  The proto-neutron star
masses for the core-collapse models are in the range 0.8--1.0\,M\sun,
and the bar-modes are limited to within this mass regime.  For the
300\,M\sun\/ case, the bar modes must develop within
70--90\,M\sun. The strain is calculated assuming a distance of 10\,Mpc
for the 15\,M\sun\/ collapse simulations, and $z=10$ for the
300\,M\sun\/ simulations.}
\end{figure}
\clearpage

\begin{figure}
\plotfiddle{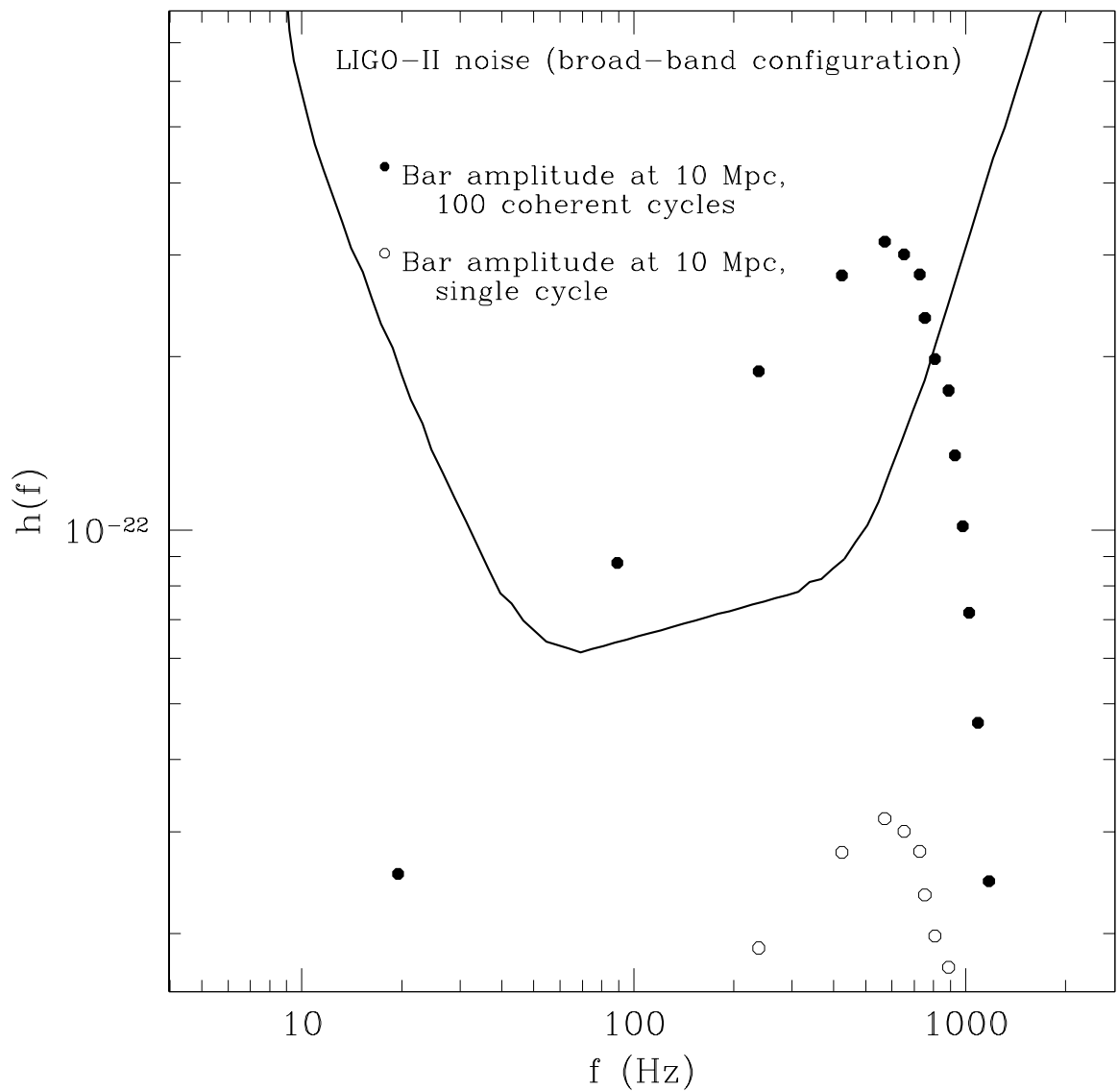}{6in}{0}{80}{80}{-240}{-100}
\caption{
\label{fig:barmode_15}
A range of possible bar mode waves emitted in the collapse of a
15\,M\sun\/ star.  We assume that all mass inside a given radius
participates in the instability and forms a bar, conserving angular
momentum as it forms.  Each point represents the waves given off from
a particular choice of radius, moving to larger radii from right to
left.  An open circle is the strain from a single rotation cycle of
the bar; a closed circle is the integrated strain that would be
measured if the bar were to remain coherent for 100 GW cycles.  The
range between the open and closed circles suggests that bar mode waves
could be of interesting strength provided they remain coherent for
a minimum of $\sim 50$--$100$ cycles.}
\end{figure}

\begin{figure}
\plotfiddle{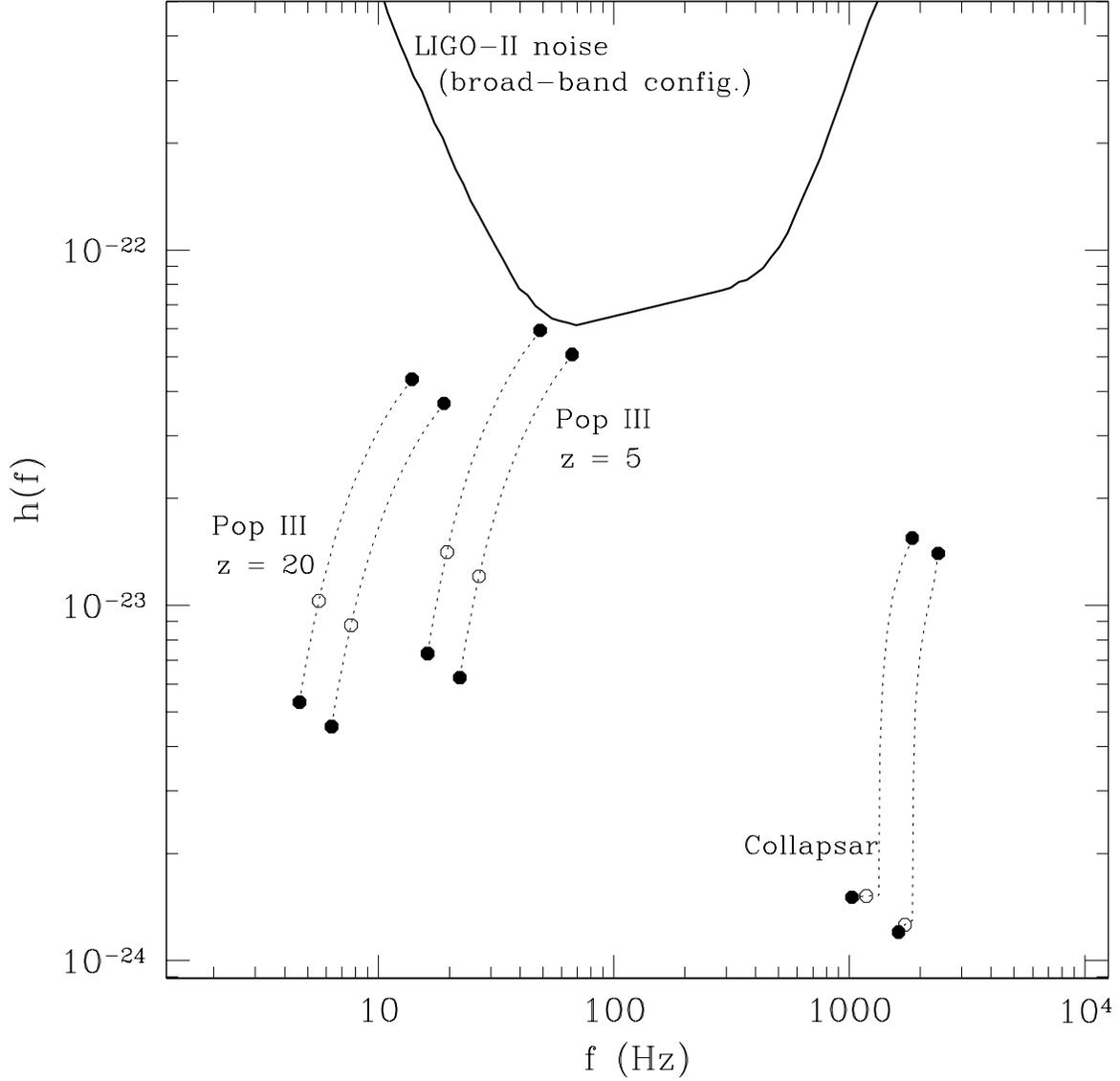}{6in}{0}{80}{80}{-240}{-100}
\caption{
\label{fig:ring}
Comparison of the characteristic wavestrain for black hole ringing to
the mean noise in enhanced LIGO interferometers (broad-band
configuration).  The two tracks for each source show the change in
strain as the parameter $\alpha_{\rm ring}$ [which apportions signal
power between $m = 0$ and $m = 2$ modes; cf Eq.\ (\ref{eq:variation})]
varies from 0 to 1.  The tracks begin at the upper right dark circles
and evolve downwards and to the left.  The open circles indicate the
half way point in time of the evolution.  Notice that the wavestrain
associated with collapsars quickly falls to its minimum value;
this is because the accretion starts out strong but is quickly
reduced.  The Population III waves are computed assuming the energy
emission parameter $\varepsilon = 0.1$ [an optimistic choice, but not
excessively so; cf.\ Eq.\ (\ref{eq:deltaE})] and $T_{\rm thump} = 0.1$
seconds (indicating a fairly clumpy flow).  The collapsar track
assumes that $\varepsilon = 0.5$ (an extremely optimistic choice) and
$T_{\rm thump} = 0.5$ seconds (indicating a very clumpy flow).}
\end{figure}

\begin{figure}
\plotfiddle{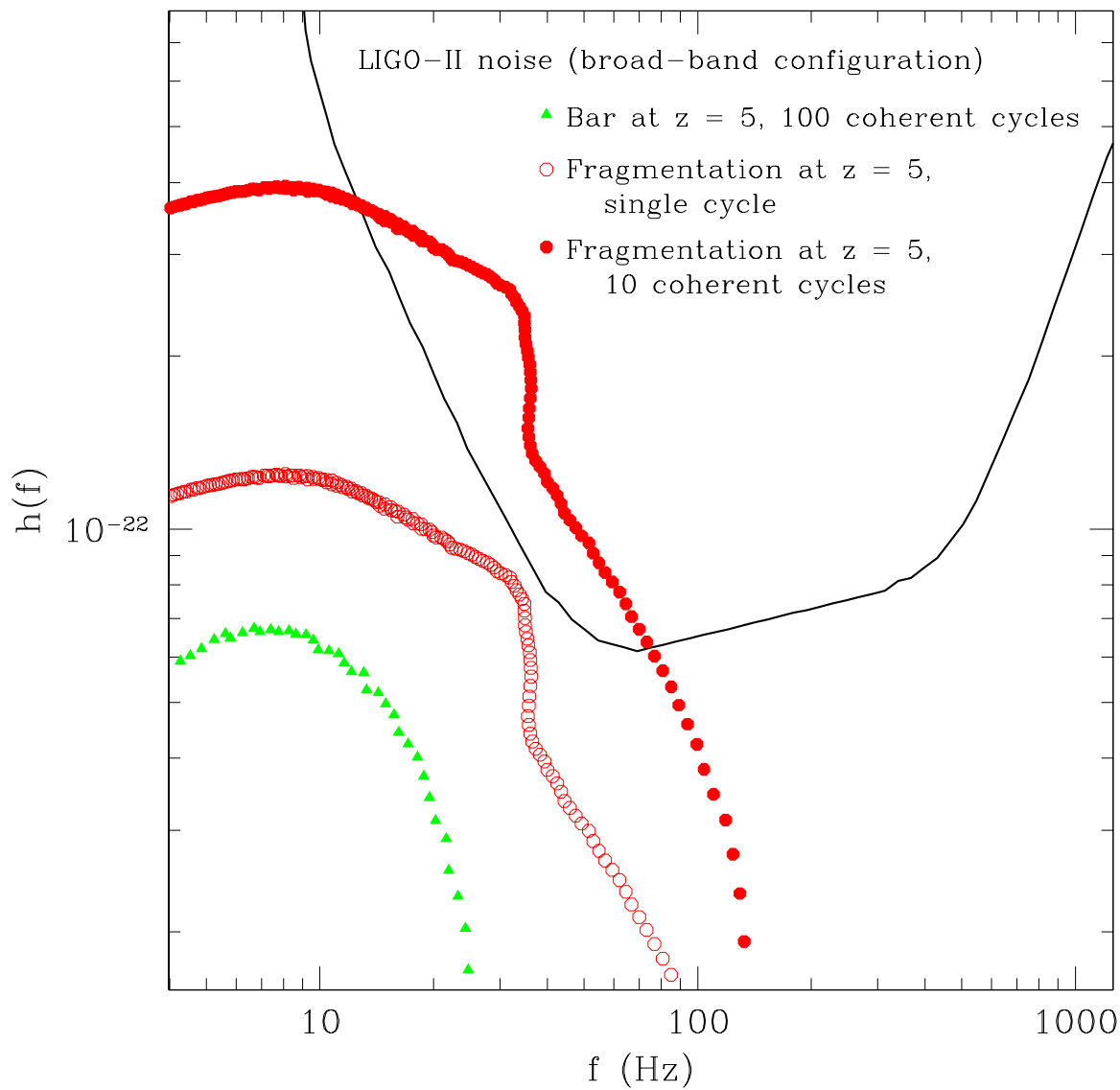}{6in}{0}{80}{80}{-240}{-100}
\caption{
\label{fig:barmode_300}
A range of possible bar mode waves emitted in the collapse of a
300\,M\sun\/ star.  Each point is calculated in the same manner as the
points given in Fig.\ {\ref{fig:barmode_15}}.  The prospects for
detecting bar mode waves are quite poor: the cosmological redshift
slides the emission frequency out of the LIGO band.  Even coherent
integration for 100 cycles is insufficient for good detection
prospects.  The waves emitted from a fragmentation instability are
more interesting: their strains and frequencies are quite a bit
higher, and may be accessible to LIGO.  If the waves from such a
fragmentation were to remain coherent for some number of cycles, they
could make an interesting source.}
\end{figure}

\begin{figure}
\plotfiddle{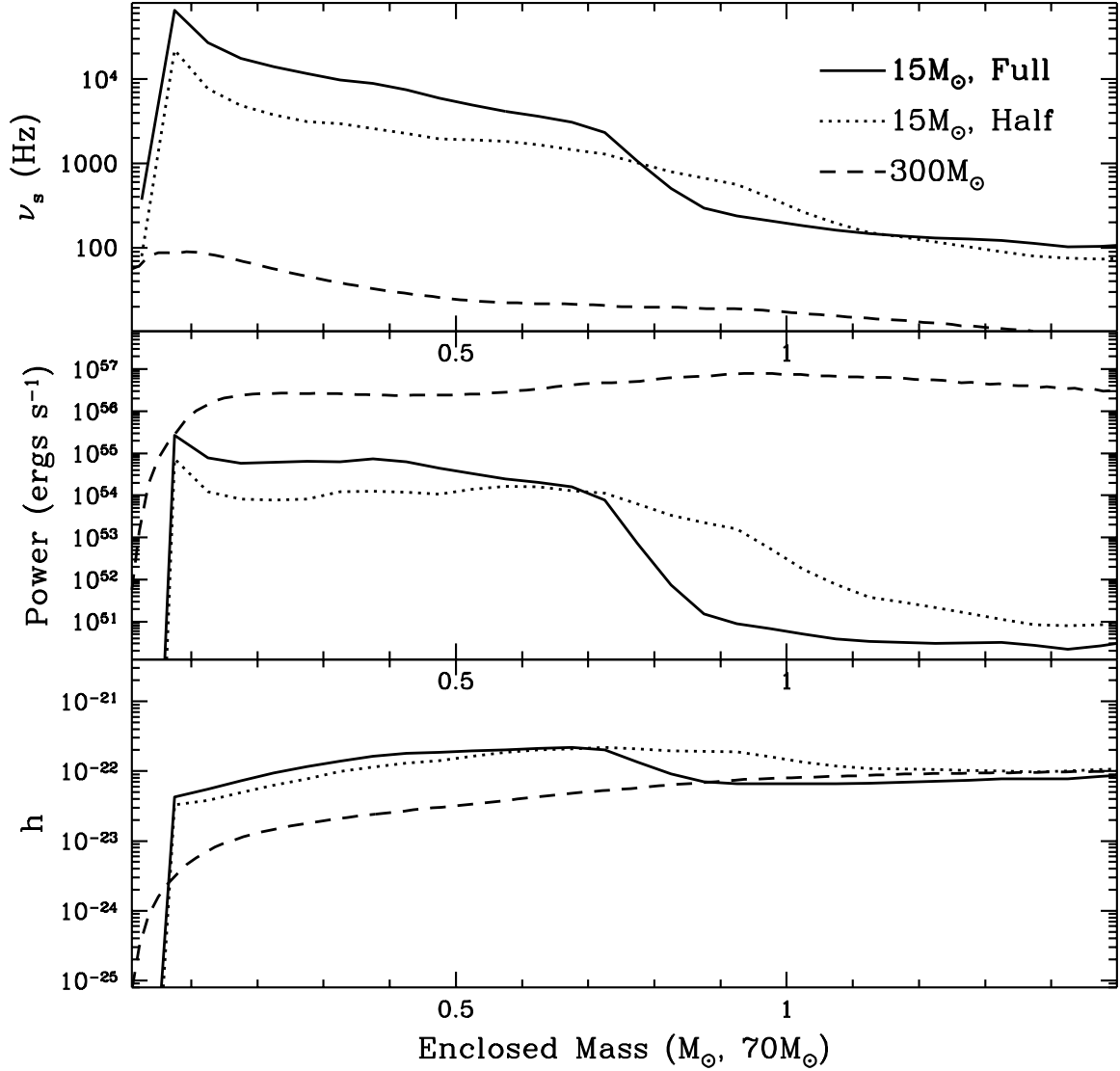}{6in}{0}{80}{80}{-240}{-100}
\caption{Binary spin period, power, and strain versus the mass which
is converted into binaries.  It is unlikely that the mass will indeed
fragment into binaries, so these calculations provide a reasonably
secure upper limit for the GW emission from stellar collapse.  The
strain is calculated assuming a distance of 10\,Mpc
for the 15\,M\sun\/ collapse simulations, and $z=10$ for the
300\,M\sun\/ simulations.}
\end{figure}
\clearpage

\begin{figure}
\plotfiddle{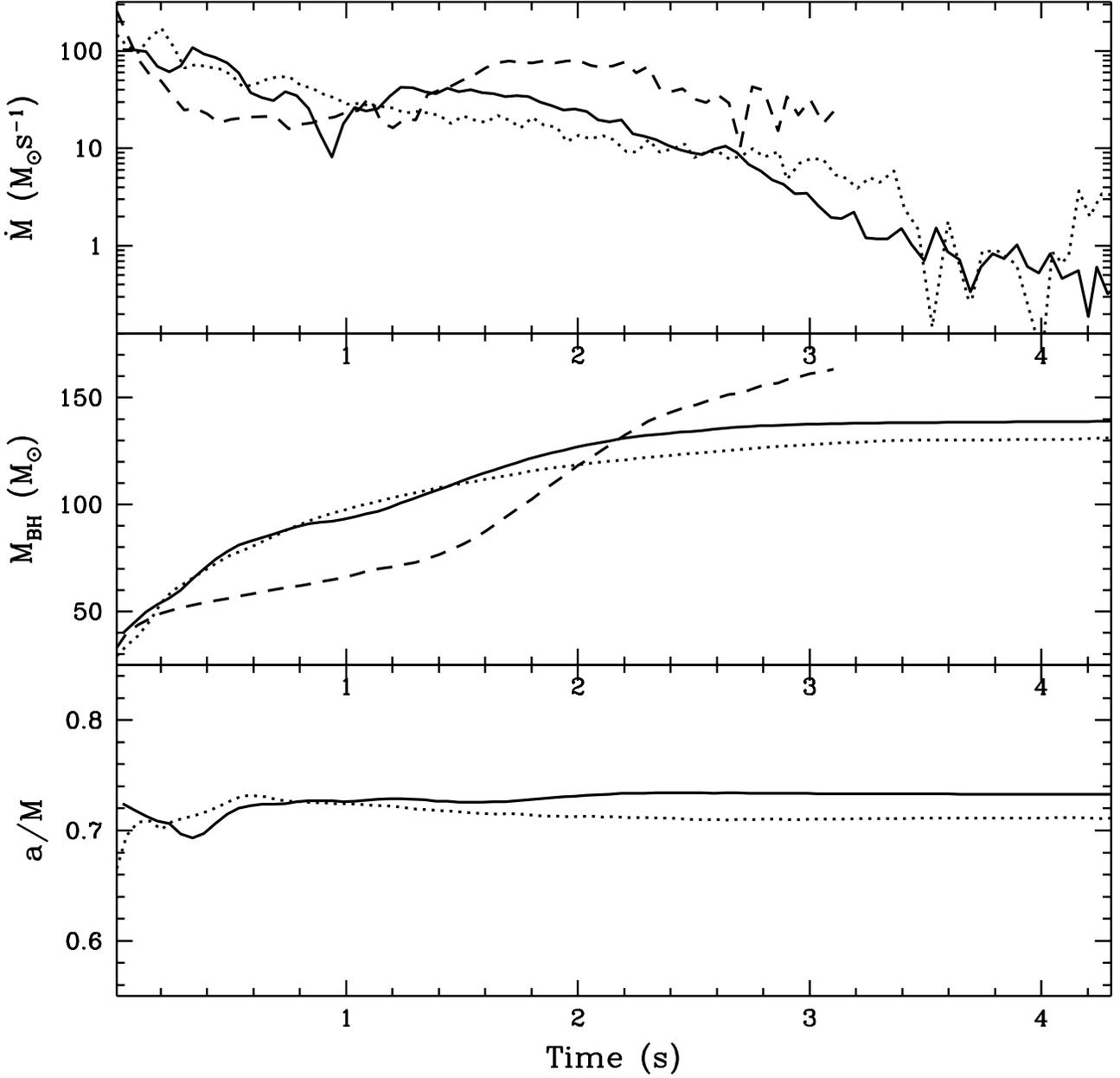}{7in}{0}{90}{90}{-280}{-100}
\caption{Accretion rate, black hole mass, and black hole spin versus
time after black hole formation in the collapse of a 300\,M\sun\/
star.  The dashed line assumes no angular momentum, and the dotted and
solid lines assume the angular momentum derived in stellar evolution
models (see Fryer et al. 2001) with and without angular momentum
transport, respectively.  The strain is calculated assuming a distance
of 10\,Mpc for the 15\,M\sun\/ collapse simulations, and $z=10$ for
the 300\,M\sun\/ simulation.}
\end{figure}
\clearpage

\end{document}